\documentclass[aps,prl,twocolumn,showpacs,superscriptaddress,groupedaddress]{revtex4}  
\usepackage{graphicx}  
\usepackage{dcolumn}   
\usepackage{bm}        
\usepackage{amssymb}   
\usepackage{epstopdf}

\usepackage{color}
\usepackage{amssymb}
\usepackage{amsmath}
\usepackage{graphicx}
\usepackage{mathrsfs}

\usepackage{float}
\usepackage[caption = false]{subfig}

\graphicspath{ {./Figures/} }
\newcommand{\abs}[1]{ \left| #1 \right| }

\begin{document}


\hspace{5.2in} \mbox{}

\title{Disentangling defects and sound modes in disordered solids}
\author{S.~Wijtmans}
\author{M.L.~Manning}
\affiliation{Syracuse University, Syracuse, New York 13244, USA }
                       .      
\date{\today}

\begin{abstract}

We develop a new method to isolate localized defects from extended vibrational modes in disordered solids.  This method augments particle interactions with an artificial potential that acts as a high-pass filter: it preserves small-scale structures while pushing extended vibrational modes to higher frequencies.  The low-frequency modes that remain are ``bare" defects; they are exponentially localized without the quadrupolar tails associated with elastic interactions. We demonstrate that these localized excitations are excellent predictors of plastic rearrangements in the solid. We characterize several of the properties of these defects that appear in mesoscopic theory of plasticity, including their distribution of energy barriers, number density, and size, which is a first step in testing and revising continuum models for plasticity in disordered solids.

\end{abstract}

\maketitle



\section{Introduction}

Under applied stress, solids can flow plastically.  In crystals, this flow is controlled via rearrangements at lattice defects, which are the dislocations \cite{Taylor1934}, easily identified by looking at the bond orientational order. In disordered solids, such as granular materials \cite{Slotterback2012,Henann2014,Coulais2014,Morgan1999} and bulk metallic glasses \cite{Su2012, Lerner2013, Schall2007} the structural defects are not easily identifiable using structural data, though the rearrangements are still localized \cite{Falk1998,Schall2007}. 

Localized defects can self-organize, changing the bulk properties of materials.  For example, shear bands, which are regions that flow faster than the rest of the material, Are thought to develop when defects co-localize ~\cite{Manning2007, Martens2012}. In other regimes, self-organization of defects leads to avalanching, where a material deforms elastically until a stress-lowering rearrangement at one defect triggers a cascade of rearrangements at other defects~\cite{Salerno2012}. Interesting memory effects \cite{Keim2013, Fiocco2014} may also be generated by self-organization of defects.

Defects play an important role in several mesoscopic phenomenological theories of plasticity, such as the theory of shear transformation zones (STZ) \cite{Falk1998,Falk2011}, which given a population and energy scale for defects, predicts how the solid will fail, and the theory of soft glassy rheology (SGR) \cite{Sollich1997}, which assumes a population of yielding mesoscopic regions.

Isolating and quantifying features of defects in disordered solids has proved difficult. Finding the precise set of particle displacements that allow the system to rearrange at the lowest energy cost is computationally difficult. Observations that the low-frequency linear vibrational modes are strongly correlated with particle rearrangements and plasticity~\cite{Widmer-Cooper2008, Tanguy2010} suggest that low-frequency localized excitations have very low energy barriers and therefore may identify defects. However, this conjecture is difficult to test because individual low-frequency normal modes are quasi-localized with long-ranged quadrupolar tails required to satisfy long-range elasticity.

One approach to overcoming this difficulty is to identify localized regions with the largest displacements in the lowest frequency modes. One finds that they do cluster into soft spots that identify locations where rearrangements are likely to occur~\cite{Manning2011}. This method has been extended to glasses at various temperatures \cite{Riggleman2014}, in different geometries \cite{Sussman2014}, and to identify the approximate directions of particle displacements \cite{Schoenholz2014}. Building on these observations, a new machine learning method has been developed to identify softness fields in disordered solids~\cite{Cubuk2015}. 
 A related approach characterizes the nonlinear response, and uses an iterative approach to identify soft nonlinear modes and their associated energy barriers\cite{Gartner2016}.

 A different, yet complementary approach analyzes the response of circular regions of particles to applied shear in many directions~\cite{Patinet2016}, and also finds evidence for pre-existing structural defects.  This method effectively predicts rearrangement sites, orientations, and energy barriers.

However, there are drawbacks to each of these approaches. The soft spots algorithm contains systematic errors due to hybridization of soft spots and sound modes~\cite{Manning2011}, while the machine learning algorithm does not identify directions of particle displacement and does not yet provide strong physical insight~\cite{Cubuk2015}. The method of Patinet et. al \cite{Patinet2016} allows for the computation of energy barriers, but it is computationally expensive and  not clearly related to the microstructure or linear response.

 In this manuscript, we develop a new method that pushes phonon-like modes out of the low-frequency spectrum, leaving isolated excitations behind.  We find that these localized excitations are excellent predictors of future rearrangements. We also characterize the number density of defects, finding that fewer soft spots are required to make accurate predictions than previously thought.

 It also allows us to characterize properties of localized excitations that are usually treated as fit parameters in mesoscopic models, including  defect size and the magnitude of associated energy barriers.  We find a localization length for defects of approximately $7.3$ particle diameters, corresponding to a soft spot containing about 150 particles.  To calculate energy barriers, we do not follow existing methods that use the adjacency matrix associated with the contact network to identify saddle points~\cite{Xu2010}as these can generate false positives~\cite{Morse2017, Zheng2016}.We develop a more robust method and find that defects have lower energy barriers than those for normal modes.
 
 \section{Simulation Model}
 
 We study 50:50 mixtures of 2500 bidisperse disks with diameter ratio 1.4, at a packing fraction of $\phi = 0.9$, significantly larger than the jamming transition at $\phi_J \simeq 0.84$~\cite{OHern2003}. Their interaction potential is Hertzian: $V= \tfrac{2}{5} \kappa \delta^{5/2}$, where $\delta$ is the particle overlap. Lengths are in units of the mean particle diameter, and energies are in units of $\kappa$. We analyze mechanically stable packings in a periodic box of linear size $L \simeq 50$ generated by an infinite temperature quench~\cite{OHern2003}.  The system is sheared using Lees-Edwards boundary conditions \cite{Lees1972}.

To find vibrational modes, we use ARPACK~\cite{ARPACK}
to calculate eigenvalues, $\lambda$, and eigenvectors, $\mathbf{e}_i$, of the dynamical matrix, $M$, which describes the linear response of the packing to particle displacements \cite{Ashcroft1976}.  At large displacements, broken contacts disrupt the linear response~\cite{Schreck2011}, but there is always a well-defined linear regime ~\cite{Goodrich2014a, VanDeen2014}. In our Hertzian packings at $\phi=0.9$, we are well within this regime.

To penalize long-range collective motion, we augment the system with an artificial square grid (lattice constant $a$) of spring-like interactions. These spring-like interactions link the coarse-grained particle motions $\widetilde{u}$, and generate an augmented potential energy $\widetilde{U}$:
\begin{multline}
\tilde{U}=\frac{1}{2}\left(\overset{n}{\sum _i}\overset{n}{\sum _j}\overset{x,y}{\sum _{\alpha }}\overset{x,y}{\sum _{\beta }}u_{i \alpha}M_{i \alpha j \beta}u_{j \beta}+
\right.\\ \left.
\overset{g^2}{\sum _k}\overset{g^2}{\sum _l}\overset{x,y}{\sum _{\gamma }}K_{kl}\left(\tilde{u}_{k \gamma}-\tilde{u}_{l \gamma}\right)^2\right),
\label{en_start}
\end{multline}

where $i,j$ are particle indices, $k,l$ are grid point indices,  Greek indices represent spatial dimensions, $g=L/a$ is the number of grid points per side, and $\widetilde{u}_{k \gamma}$ is a Gaussian weighting of particle displacements and represents an effective grid point motion
 as described in the Appendix  and illustrated schematically in Fig. \ref{fig_1}. With this augmented energy $\widetilde{U} = U + U^{\dagger}$, the dynamical matrix is $ \widetilde{M} = M + M^{\dagger}$ with

\begin{multline}
M^{\dagger }=\overset{g^2}{\sum _k}\overset{g^2}{\sum _l}K_{kl}\delta _{\alpha \beta }\left(W(i,k)W(j,k)\right.\\\left.-2W(i,k)W(j,l)+W(i,l)W(j,l)\right),
\label{M_dagger}
\end{multline}
and 
\begin{equation}
W(i,l)=\text{Exp}\left[-{\sum _{\eta }^{x,y}}\left(x_{i \eta }-l_{\eta }a\right)^2/\sigma^2\right].
\end{equation}

We refer to modes associated with $U$ as standard modes and those associated with $\tilde{U}$ as augmented modes. As we have implemented the effective spring network on a square grid, $K_{kl} = K \delta_{k,l\pm 1}$. The ``motion'' of each grid point ($\tilde{u}$) is illustrated schematically in Fig. \ref{fig_1}(a); average displacements in the same direction are penalized.
\begin{figure}
\includegraphics[width=3.4in]{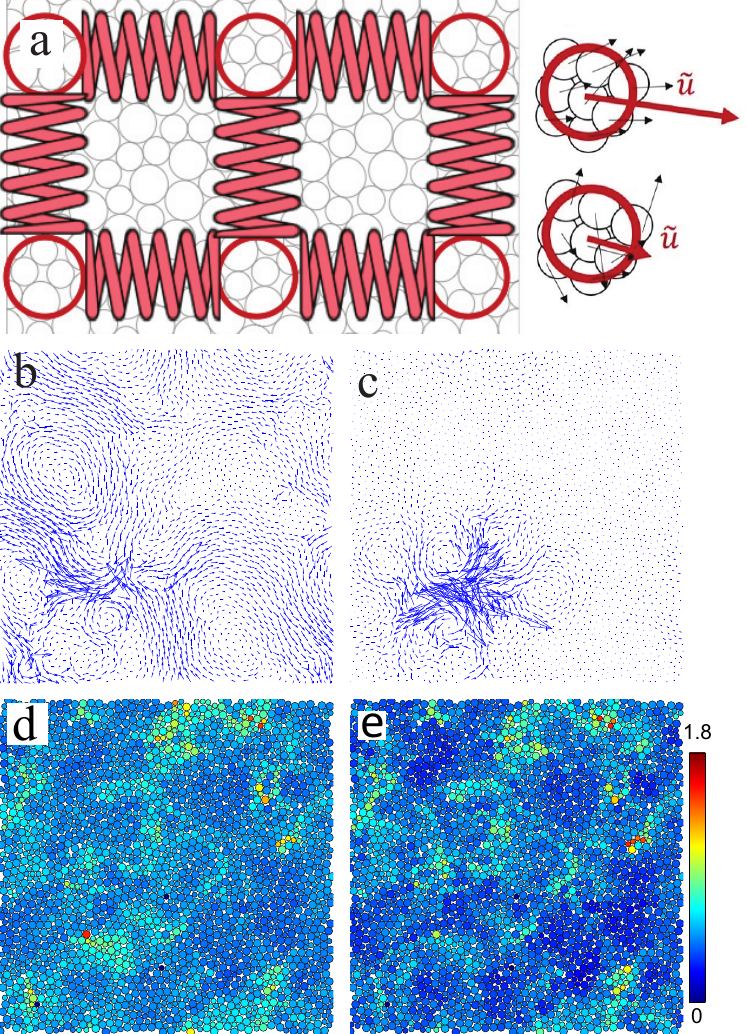}
\caption{(color online). (a) A square grid of points is connected by a spring-like interaction. Movement of a grid point is defined by the weighted particle displacements. (b) A typical hybridized vibrational mode from a standard dynamical matrix. (c) A typical localized mode from an augmented dynamical matrix. (d,e) Sum of the magnitudes of particle displacements in the 30 standard (d) and augmented (e) modes with the lowest frequencies. The colorbar represents the magnitude of the summed polarization vector.
}
\label{fig_1}
\end{figure}

There are three parameters for the augmentation: the width of the Gaussian weighting for each gridpoint $\sigma$,  the spacing between grid points $a$, and the strength of the augmentation $K$. We do not want the augmented energy to affect the energy of individual particles. As discussed in the appendix, the sum of Gaussians is nearly unity when $\sigma$ is equal to or larger than $a$; therefore we choose $\sigma=a$ for the remainder of this manuscript.

\begin{figure}
\includegraphics[width=3.5in]{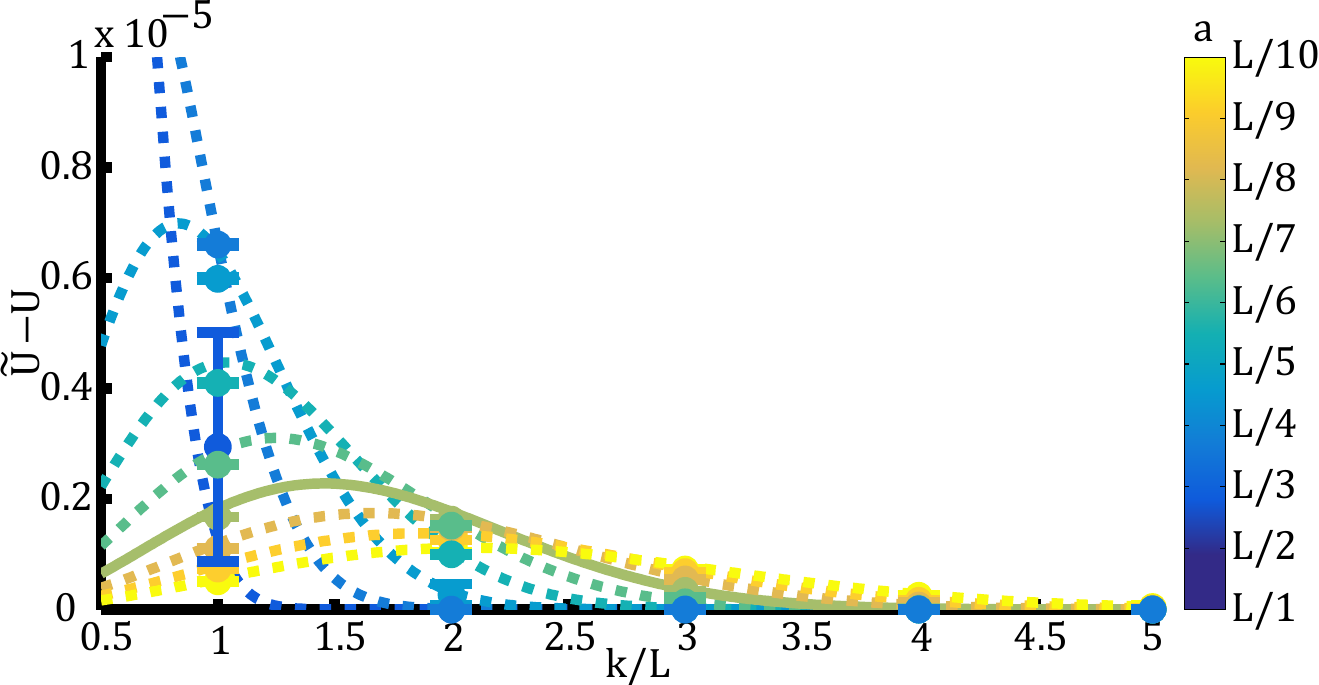}
\caption{ (color online). Energy due to augmentation, $\widetilde{U}-U$, of a plane wave of wavenumber $k$ for various number of grid points spacing $a$ ranging from $L/2$ to $L/10$ at  $L=46$.  The solid line corresponds to $a=L/7$. Results are shown both for the analytic prediction, as lines, and numerical results, as points, which do not differ significantly.
%
}
\label{fig_param}
\end{figure}

We want to choose $a$ such that the augmented energy  acts as a high-pass filter, increasing the energy of modes with small wavenumbers.  Therefore, we compare the standard ($U$) and augmented ($\widetilde{U}$) energies associated with a field of displacements that vary in amplitude as a plane wave with wavenumber $k$. An analytic expression for this quantity is derived in Appendix A.  The analytic and numerical results, shown in Fig. \ref{fig_param}(a), suggest that wavevectors smaller than $\sim L/2a$ are significantly penalized, and that there is a systematic decrease in sensitivity as $a$ decreases. To balance these effects, we choose  $a = L/7$  for the remainder of this manuscript.  

A more subtle question is how to balance the magnitude of the augmented energy and the standard energy, which is controlled by the parameter $K$. We want the augmented energy to be sufficient to push out plane waves, without affecting finer scale structure. To do so, we examine the number of well-localized modes as as function of $K$,  as described in Appendix B. We find that $K=0.01$
is sufficient to penalize plane waves without altering the fine-scale structure.



\section{Results}

Having specified reasonable values for $a$ and $K$, we now explore the eigenspectrum of the augmented system. Fig. \ref{fig_1}(b) shows the particle displacements in a typical low-frequency hybridized mode derived from the standard dynamical matrix $M$.  A typical low-frequency eigenvector of the augmented dynamical matrix $\widetilde{M}$, shown in Fig. \ref{fig_1}(c), is localized. The inset shows an exponential decay in amplitude away from the center of the defect, highlighting the absence of long-range quadrupolar tails~\cite{Maloney2006}.

 Fig.~\ref{fig_1}(d,e) show the sum of the magnitudes of the 30 lowest frequency modes for typical standard and augmented matrices, respectively.  In Fig.~\ref{fig_1}(e), large magnitudes occur in the same localized regions as in Fig.~\ref{fig_1}(d), indicating that the augmented potential does not interfere with small-scale structure or alter the locations of the soft spots. The augmentation does suppress the background associated with extended excitations, making soft spots easier to identify.

With this choice of parameters we can measure localization with the radius of gyration~\cite{Donati1997}, given by
\begin{equation}
R_g=\sqrt{\frac{\sum_{i=1}^N ( \abs{r_i-r_{cm}}^2 \abs{\mathbf{e}_i})}{\sum_{i=1}^N (  \abs{\mathbf{e}_i}) }}.
\end{equation}  

\begin{figure}
\includegraphics[width=3.5in]{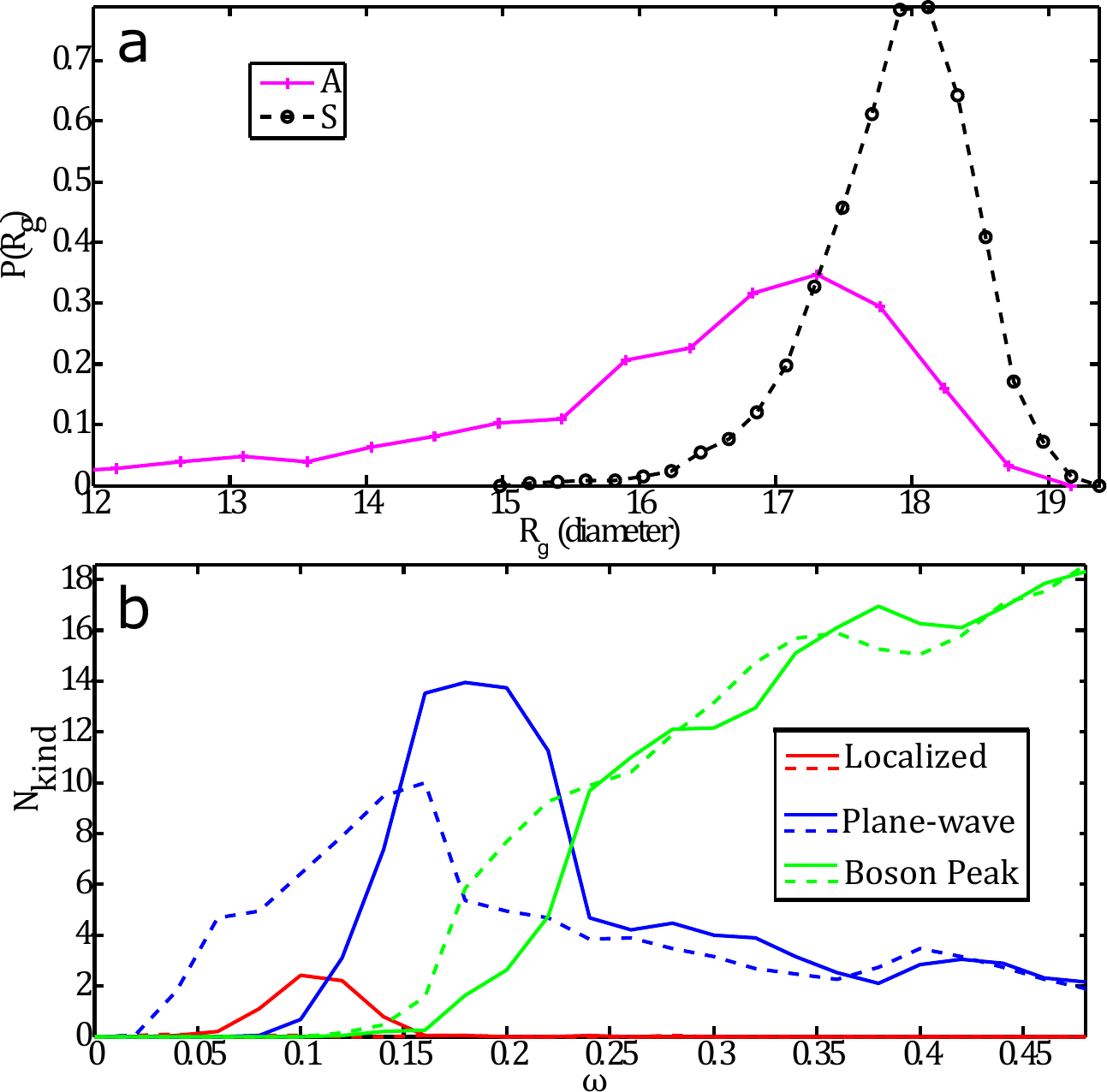}
\caption{ (color online).  
a) The radius of gyration of the 20 lowest frequency modes in 100 realizations are much smaller for augmented modes (solid line) than for standard modes (dashed line).
b) Density of states plotted as number of modes per packing, for augmented (solid) and standard(dashed) modes.  Localized modes (red) only appear in the augmented DOS, while the plane waves (blue) in the augmented system have increased energy.
}
\label{fig_2}
\end{figure}

To confirm that modes with increased energies are indeed extended, we split eigenvectors into three groups. 
We find that the normal modes rarely have a $R_G \lessapprox 16$ particle diameters as shown in Fig. \ref{fig_2}a, therefore we denote modes with $R_G < 16$ as ``localized''.
Remaining modes are characterized by an integral $I_f$ over the Fourier spectrum and  $L_2$ distance $L_{BP}$ between their cumulative distribution functions (cdfs) and the typical boson peak cdf ~\cite{Manning2015b}. We find that $ I_f = 0.3-2.3 L_{BP}$ is the best separating plane in $I_f - L_{BP}$ space between ``phonon-like'' and ``boson peak'' modes.

Fig. \ref{fig_2}(b) shows that the augmented potential is altering the spectrum as intended.  The augmented eigenpectrum shown by solid lines contains a significant number of localized modes, and plane waves are pushed to higher frequencies compared to standard modes (shown by dashed lines).


\subsection{Size and number density}

In mesoscopic models for plasticity, two important parameters are the number density and size of localized excitations, and therefore one of our goals is to extract distributions for these parameters directly from simulations.

\begin{figure}
\includegraphics[width=3.5in]{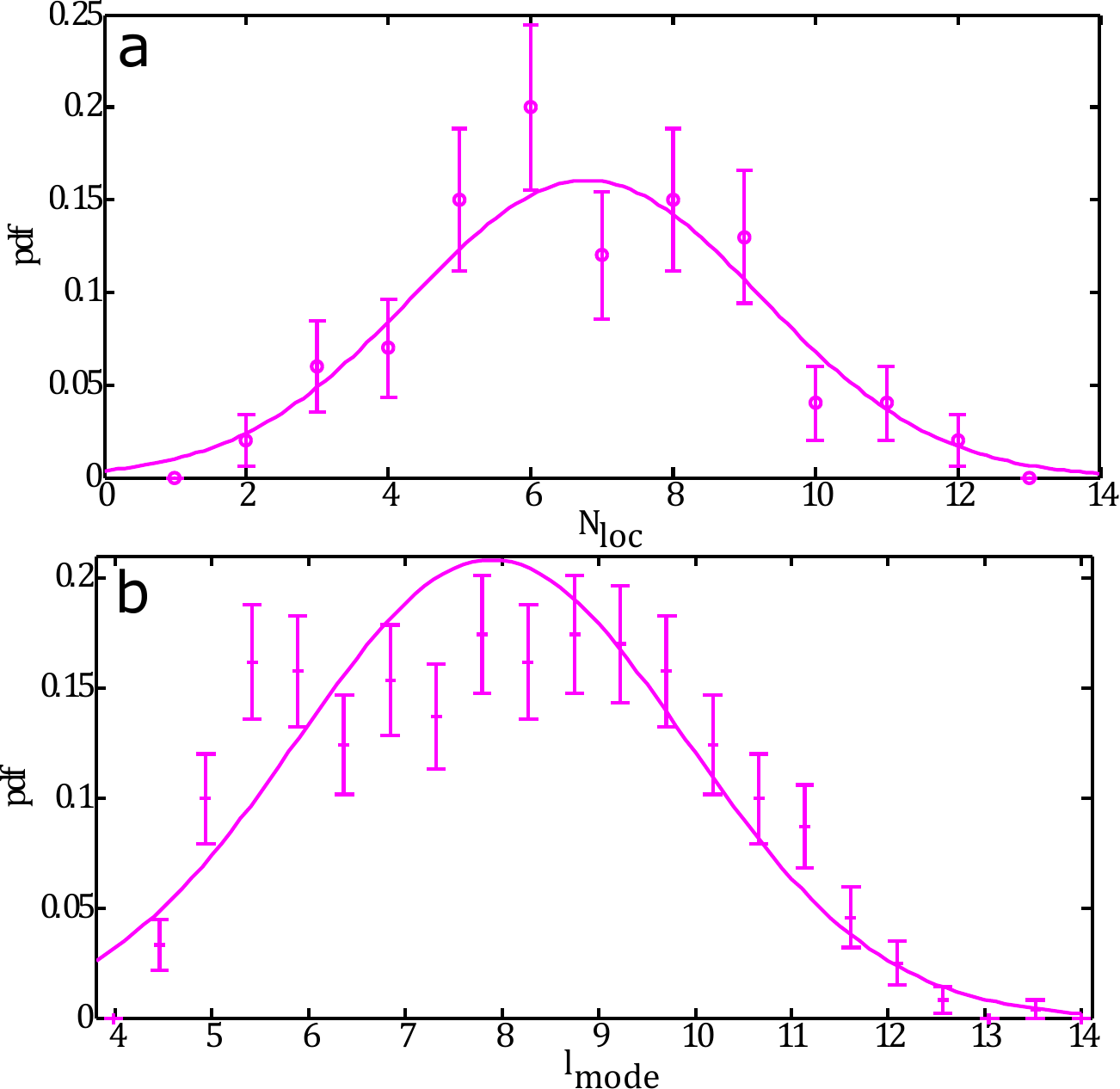}
\caption{a) The number of modes with $R_G < 16$ in the augmented system. The line is a Gaussian fit with $\mu=6.79$ and $\sigma=2.45$.
b) The length scale extracted from an exponential fit to the radially averages distribution of displacement vectors for augmented localized modes. The line is a Gaussian fit with $\mu=7.89$ and $\sigma=2.01$.
}
\label{fig_num_loc}
\end{figure}

The number of localized augmented modes should serve as a lower bound on the number of localized defects. As seen in Fig. \ref{fig_param}, our choice of grid spacings suppresses wavenumbers of up to $ k\cong 3$ so localized excitations with frequencies above that range will not be isolated. The distribution for the number of localized excitations in our 2D box of 2500 particles is shown in Fig.~\ref{fig_num_loc}a -- the average is about 7.

We can also extract a length scale for each localized augmented mode by fitting an exponential function to the decay in displacement amplitude away from the defect center. The distribution of lengthscales associated with these exponential fits is shown in Fig \ref{fig_num_loc}b, with a mean of $7.89$ particle diameters. We have checked that this length is independent of our choice of grid spacing $a$.

To see how these results compare to previous approaches, we characterize the size and number density of the localized excitations using the soft spots approach~\cite{Manning2011}. This method identifies clusters of soft particles by grouping together the $N_p$ particles with the largest motion from the lowest $N_m$ frequency modes, and optimizes $N_p$ and $N_m$ in order to maximize the correlation $C_{sr}$ of spots with rearrangements. $C_{sr}$ is a very strict criterion comparing the particles that have large displacements in soft spots and rearrangements, respectively. 

For standard normal modes at a packing fraction of $0.9$, the best correlation with rearrangements is found when we include the $N_m = 25$ lowest frequency modes and the $N_p = 20$ particles in each mode.  This corresponds to $14 \pm 3$ soft spots in each packing of $2500$ particles, with an average soft spot size of $ 14.3 \pm 5.7$ particles.

These results are quite different from those we found with our new augmented approach -- the augmented approach finds about half as many localized excitations in the same packing ($7$ compared to $14$) and excitations that are an order of magnitude larger (a lengthscale corresponding to $150$ particles compared to $14$ particles). Typical defect sizes reported in the literature vary widely and are typically between these two values~\cite{Schall2007, Falk1998, Shavit2013, Ding2014, Lerner2009, Schrder2000}.

\begin{figure}
\includegraphics[width=3in]{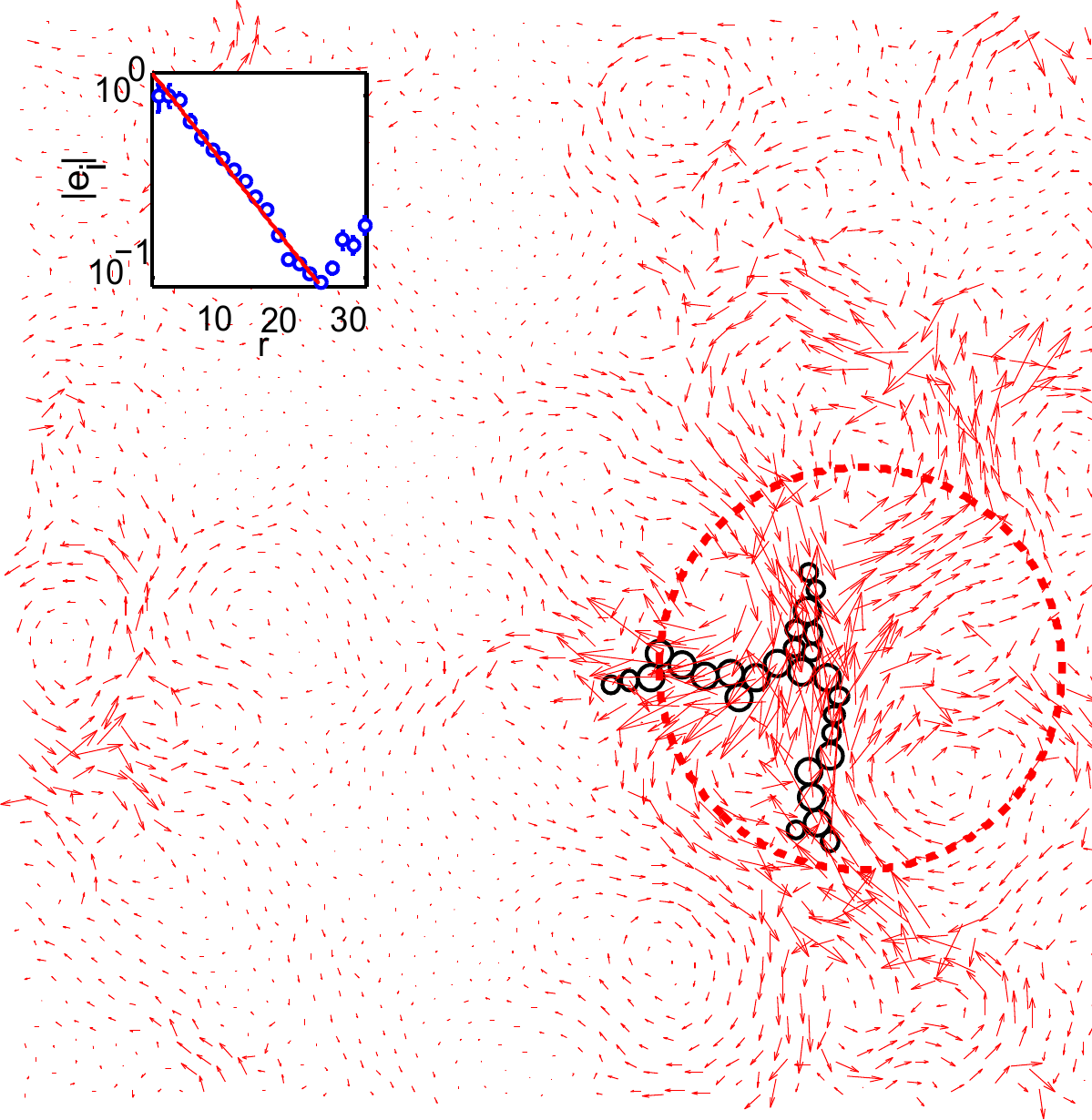}
\caption{ (color online). A soft cluster compared to the most similar augmented mode, of $R_G=15.3$. The localization length of the augmented mode ($l_{mode}=8.8$) is represented as a circle.
 (inset) Blue points are the radially averaged magnitudes of particle displacements $\abs{\mathbf{e}_i}$ a distances $r$ from largest particle displacement; the red line is the best exponential fit with decay length $l_{mode} = 8.8$.
}
\label{fig_single_cluster}
\end{figure}

To unpack this discrepancy, we first look more closely at the size of localized excitations. Fig~\ref{fig_single_cluster} compares displacements in a localized augmented mode to the locations of particles identified by the soft spot algorithm. The soft spot singles out the particles with very large motion during a rearrangement, while the augmented mode retains smaller displacement vectors. Because the defect core is far from compact (perhaps even with string-like structures reported previously~\cite{Donati1998}), the spatial extent of the soft spot and the localization length (derived from an exponential fit to the displacement field) are actually very similar. This suggests that the {\emph size} of a localized excitation may depend quite a bit on how size is measured.

Next, we investigate the number density of localized excitations.  A possible weakness of existing soft spots algorithms is that they are only able to demonstrate that a given number of soft spots is sufficient for predicting plasticity, but not that the number is necessary -- e.g. these methods may be identifying more spots than are needed to predict rearrangements. 

To test this hypothesis, we run the soft spot analysis using only the \emph{ localized} augmented modes.  As expected, this generates about the same number of soft spots as localized modes (an average of $7$ spots). The results are largely insensitive to the the number of particles per mode, $N_p$, for $N_p$ between $20$ and $50$, as we might have expected from our analysis of soft spot size above. We report results for $N_p =25$ so that the average soft spot size matches that for standard modes.

An interesting questions is whether the $7$ spots identified by our new augmented method are just as effective at predicting plasticity as the $14$ spots generated by the old method.  Interestingly, the overall average correlation is nearly identical $C_{sr} = 0.777$ for standard modes and $C_{sr} = 0.781$ for localized augmented modes. Fig~\ref{fig_mode_dot}(a) shows $C_{sr}$ for standard modes and only localized augmented modes as a function of $\gamma-\gamma_c$, which is the strain required to activate the next rearrangement.  We see that close to the event the $7$ localized augmented modes are slightly better than the $14$ standard soft spots at predicting rearrangements, and nearly as good even far from the event. This suggests that our augmented potential is identifying a subset of localized excitations that predict plasticity, and that the number density of such excitations is actually significantly smaller than previously thought.





\subsection{Direction information and energy barriers}

\begin{figure}
\includegraphics[width=3.5in]{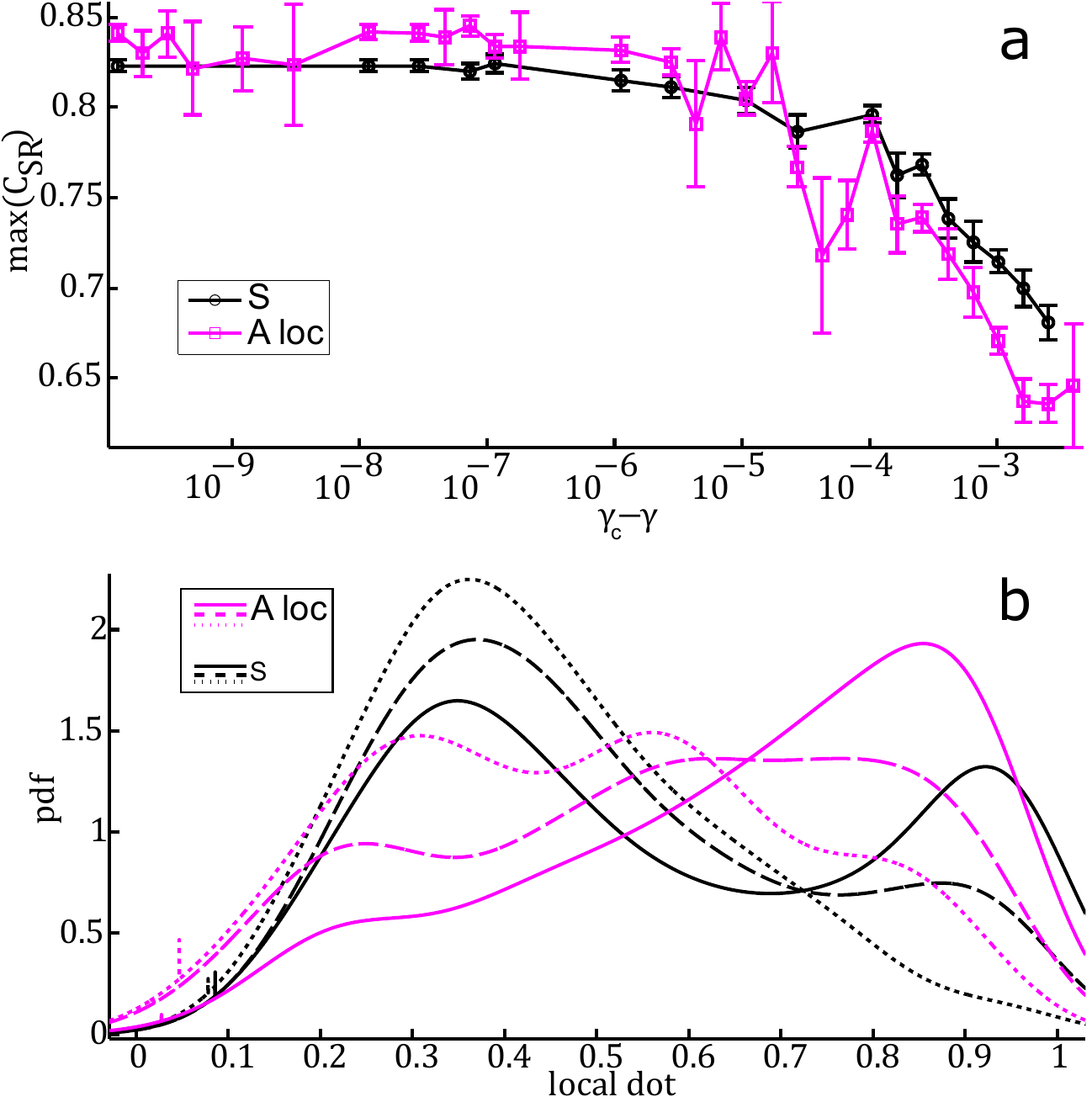}
  \caption{a)  Correlation function $C_{SR}$  between soft spots and rearrangments generated from the localized augmented modes (magenta) are very similar to those for standard modes (black), and more accurate far from the event. Events are binned as a function of the strain required to get to the next particle rearrangement $(\gamma_c-\gamma$).
  b)  Comparison of the highest dot product between localized ($R_G<16$) augmented modes (magenta) and standard modes (black). The various curves represent different distances in strain from the rearrangement. The solid lines correspond to strains close to a rearrangement ($(\gamma-\gamma_c)<10^{-4.5}$), the dashed lines are $10^{-4.5}\leq (\gamma - \gamma_c-)\leq 10^{-3.5}$, and the dotted line represents $(\gamma -\gamma_c)\geq 10^{-3.5}$.  The pdf was created from a cdf using a Gaussian smoothing filter of $\sigma=.07.$  We see that the localized augmented modes tend to match the rearrangement much better than the standard mode, especially far from the rearrangement as demonstrated by the magenta curves having larger weights near unity. 
  }
  \label{fig_mode_dot}
\end{figure}

Of course, one of the main benefits of this algorithm is that we can calculate not only the locations of the localized excitations but also the directions of particle motions in an excitation. To see if this directional information is also predictive for particle rearrangements, we take the scalar dot product between each of the localized augmented modes and the rearrangement, as well as the lowest-frequency standard modes. We restrict the dot product to a circular region of radius $r_c$ around the localized event, to avoid noise associated with dot products of many random vectors with small magnitudes.  We choose $r_c  = L/5$ which is slightly above the localization length, but results are not highly sensitive to the choice of $r_c$.  We also restrict ourselves to non-avalanche rearrangements, where the rearrangement has over an 80\% overlap with the mode that goes to zero at the critical strain~\cite{Manning2011}. Fig. ~\ref{fig_mode_dot} b shows the probability distribution for the highest dot product amongst all modes. There is a much higher probability of finding  a dot product near unity for the augmented modes compared to the standard modes, indicating that augmented modes are better predictors of the rearrangement dynamics, even at strains long before a rearrangement event.

\begin{figure}
\includegraphics[width=3.5in]{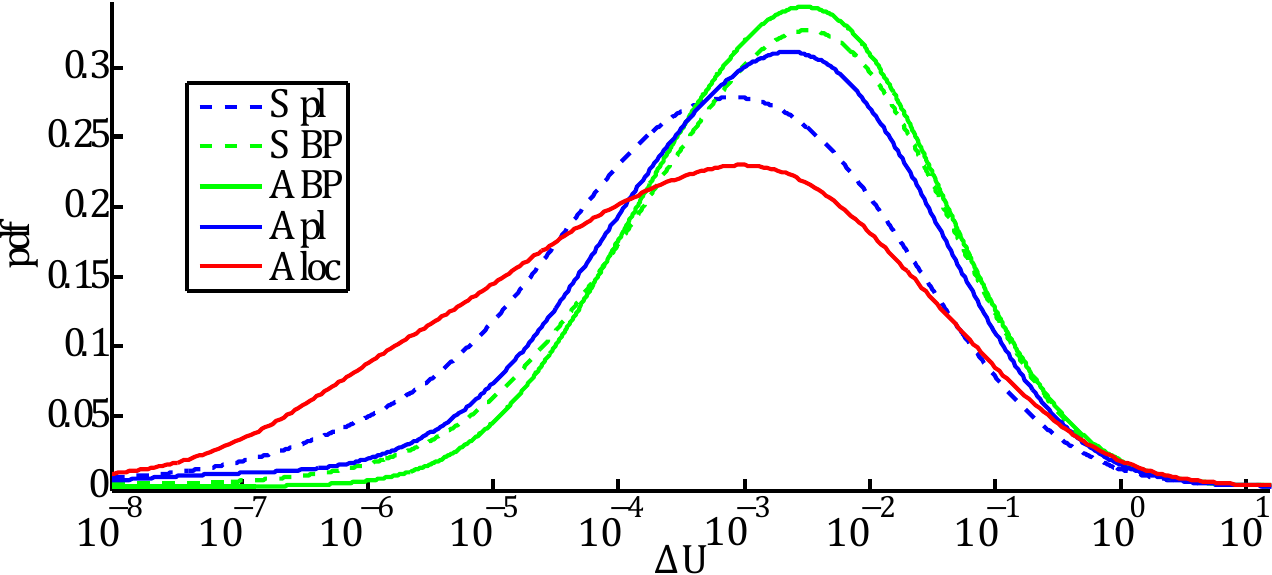}
\caption{ {\bf Energy barriers. } 
Probability distribution of energy barrier for standard (dashed), and augmented (solid) modes, collected over the  50 lowest frequency modes in 100 realizations are plotted separated by mode type as in Fig. \ref{fig_2}.  Localized augmented modes are more likely to have very low energy barriers.  Curves were produced by smoothing the cumulative distribution function with a Gaussian filter of width 0.7.
}   
\label{fig_barrier}
\end{figure}

This algorithm for the first time allows us to calculate the distribution of energy barriers associated with bare defects.  We displace particles along each eigenvector and use the LBFGS line search algorithm~\cite{Nocedal1980} to minimize the potential energy at every step. At the first step where the minimized state is different from the initial state, we identify a particle rearrangement and define an energy barrier $\Delta U$ as the difference between the initial energy and the maximum energy attained~\cite{Xu2010}.  

The definition of a particle rearrangement or new state can be subtle. Although previous studies in particulate matter have used changes in the contact network adjacency matrix to identify energy barriers~\cite{Xu2010}, not all contact changes identify saddle points in the potential energy landscape~\cite{VanDeen2014, Morse2017}.  In Appendix C, we compare several methods for identifying saddle points in the potential energy landscape, and identify a subset that are self-consistent and match our expectations for saddle points. Using one of these self-consistent definitions, we calculate the energy barriers for the 50 lowest frequency standard and augmented modes over 100 realizations, separated by mode type. Fig. \ref{fig_barrier} shows that the localized augmented modes are more likely to have very low energy barriers compared to standard modes, although the average value is similar.

\section{Discussion and Conclusions}

Using a simple physically-motivated augmented potential, we were able to push extended low-frequency vibrational modes to higher frequencies and isolate localized ``bare'' defects. We demonstrate that these localized excitations are excellent predictors of both the location and displacements associated with particle rearrangements in a disordered solid. Finally, we characterize the lengthscales, number density, and energy barriers associated with these excitations.

These results should immediately improve continuum models for plasticity in amorphous solids.  The energy barriers associated with defects in disordered solids are an important parameter in both STZ and SGR models. While these were previously fitting parameters, one can now extract them from a simulation and test specific model predictions. 


A future direction of inquiry is to study the creation, annihilation, and activation of defects in sheared simulations and compare to continuum model assumptions ~\cite{Manning2007, Cao, Keim2014, Schall2007, Perchikov2014, Chen2011}.  In particular, it would be instructive to use the directional and energy barrier data to predict which defect would activate given a particular shear direction.


We know that other methods for identifying soft spots such as those calculating local yield stress~\cite{Patinet2016}, analyzing nonlinear modes~\cite{Gartner2016b}, or using machine learning~\cite{Cubuk2014} also correlate well with particle rearrangements. Therefore, it would be useful to systematically compare these methods on the same packing and determine whether the identify the same localized excitations.

While we have used a multiple contact-change metric for determining a new state to calculate energy barriers, there are many other choices. Ongoing work~\cite{Morse2017} suggests that all saddle points are accompanied by a drop in stress, while contact changes that are not associated with saddle points do not have a substantial stress drop. We do not use stress drops as an indicator of saddle points here because they are difficult to identify along the non-physical particle paths prescribed by our augmented modes, but we expect this alternate definition to be useful in other systems, such as those under simple shear.


As this method can robustly identify defects far from rearrangements, we are now well poised to study their dynamics. For example, we hope to study the creation and annihilation of defects under shear, and whether their statistical properties (such as the lengthscale, number density, or energy barrier) are different as a function of material preparation or shearing history. This new technique should also allow us to study the spatial organization of defects in processes such as shear banding that lead to catastrophic failure.

Additionally, it would be interesting to try to extend this algorithm to systems that are not in mechanical equilibrium. For example, during an avalanche in a highly jammed packing we expect that most of the motion will be along a few floppy modes, so that most of the curvatures in the potential energy landscape are still positive and amenable to study via vibrational mode techniques. This would allow us to observe the activation and formation of defects, to understand if avalanches mostly result from one defect triggering other, pre-existing defects, or if avalanches create and activate new defects.

The self-organization of defects may also provide insight into the reversibility dynamics seen by Keim et. al \cite{Keim2014,Keim2013,Paulsen2014}.  The evolution of soft spots as a function of time and their organization in the highly reversible system acquired after many cycles might allow a mesoscopic description of memory formation in materials. 

\section{Acknowledgements}

We thank Jim Sethna, Andrea Liu, Matthias Merkel, Peter Morse, and Dapeng Bi for discussions. This work was supported in part by NSF-DMR-1352184 (MLM and SW), and the Alfred P. Sloan Foundation (MLM), and the Simons Foundation Grant No. 454947 (MLM). Computing resources were provided by Syracuse University and NSF-ACI-1541396.

\title{Augmentation Appendix}

\author{Sven Wijtmans}

\date{\today}

\maketitle


\section{\label{sec:aug_calc}Appendix A}

In this Appendix, we derive the augmented dynamical matrix that allows us to separate localized modes from extended plane-wave-like modes at low frequency in a disordered solid.  Latin indices are used to label particles and greek indices to label cartesian components. All summations are explicit.

The low frequency vibrational modes of a disordered solid contain localized excitations at defects hybridized with extended plane-wave-like modes and are the eigenvectors of the Dynamical matrix $M_{i\alpha j \beta}=\frac{\partial^2 U}{\partial u_{i\alpha} \partial u_{j\beta}}$. In order to examine the bare defects, we must separate these two types of modes. To prevent hybridization, our goal is to increase the energy of the plane-wave-like modes without increasing the energy of the localized modes.

To do this, we add an extra term to the total energy of the packing $U$ that represents a grid of virtual points connected by a spring-like interaction.  The motion of each grid point is defined as the motion of the particles near it, weighted by a Gaussian function of the distance between them. The new energy is then

\begin{multline}
\tilde{U}=\frac{1}{2}\left(\overset{n}{\sum _i}\overset{n}{\sum _j}\overset{x,y}{\sum _{\alpha }}\overset{x,y}{\sum _{\beta }}u_{i \alpha}M_{i \alpha j \beta}u_{j \beta}+
\right.\\ \left.
\overset{g^2}{\sum _k}\overset{g^2}{\sum _l}\overset{x,y}{\sum _{\gamma }}K_{kl}\left(\tilde{u}_{k \gamma}-\tilde{u}_{l \gamma}\right)^2\right),
\label{en_start_aug}
\end{multline}

where

\begin{align}
\tilde{u}_{k \gamma}&=\frac{\overset{n}{\sum _p}u_{p \gamma}\text{Exp}\left[ -\overset{x,y}{\sum _{\eta }}\left(x_{p \eta}-k_{\eta}a\right)^2/\sigma ^2\right]}{\overset{n}{\sum _p}\text{Exp}\left[ -\overset{x,y}{\sum _{\eta }}\left(x_{p \eta}-k_{\eta}a\right)^2/\sigma ^2\right]} ,
\\ 
k_x&=\text{Floor}[k/g],\\
k_y&=\text{Mod}[k,g],\\
g&=L/a,
\label{eq_u_aug}
\end{align}

where $g$ is the number of grid points per side, $a$ is the spacing between grid points, and $L$ is the side length. $K_{kl}$ sets the connectivity and the strength of the connection between grid points.  In this work, we connect adjacent grid points on a square grid, so that $K_{kl} = K \delta_{k,l\pm 1} $.

\begin{figure}
\includegraphics[width=3.4in]{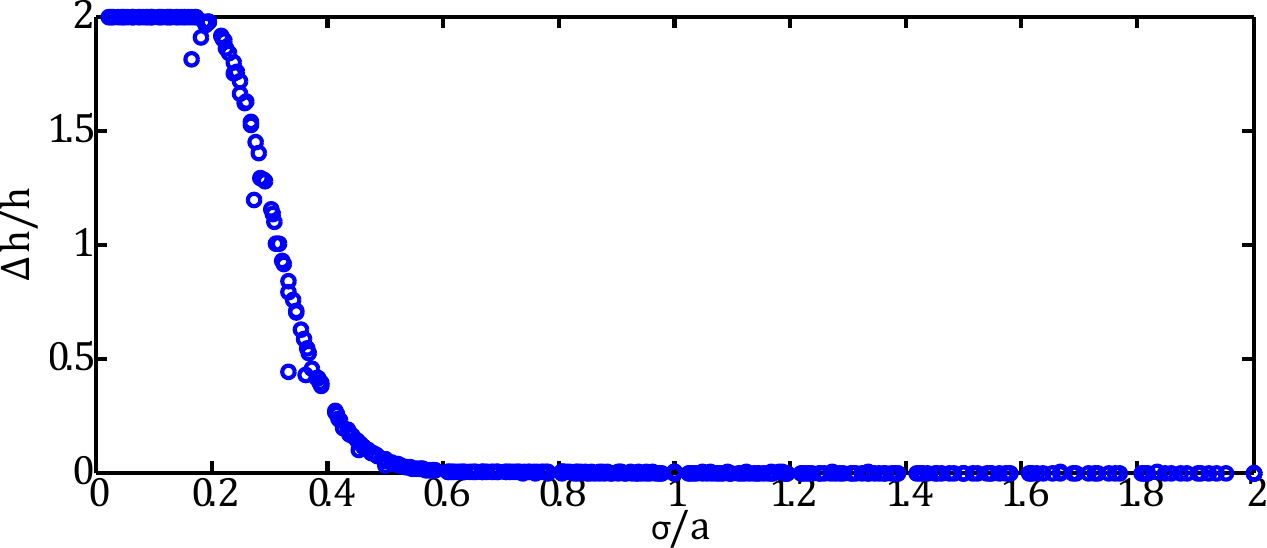}
\caption{ Ratio of height variation $\Delta h$ to mean height $\bar{h}$ of a grid of gaussians of width $\sigma$ and spacing $a$ between each peak.
}
\label{fig_gauss_sig_aug}
\end{figure}

The width of the Gaussian weighting is set equal to the grid spacing, as the sum of a grid of Gaussians with width equal to the spacing is flat to within $10^{-6}$, as seen in Fig.\ref{fig_gauss_sig_aug}.  We divide by the Gaussian contribution of each particle near the grid point to normalize by particle density.

The total augmented energy can be written as a quadratic function in terms of the standard dynamical matrix $M$ and the dynamical matrix $M^\dagger$ corresponding to the augmented potential:

\begin{equation}
\tilde{U}=\frac{1}{2}\left(\overset{n}{\sum _i}\overset{n}{\sum _j}\overset{x,y}{\sum _{\alpha }}\overset{x,y}{\sum _{\beta }}u_{i\alpha
}\left(M_{i\alpha j \beta }+M_{i\alpha j\beta }^{\dagger }\right)u_{j\beta}\right).
\end{equation}

To simplify notation, we introduce a weighing function
\begin{equation}
W(i,l)=\frac{\text{Exp}\left[-{\sum _{\eta }^{x,y}}\left(x_{i \eta }-l_{\eta }a\right)^2/\sigma^2\right]}{\overset{n}{\sum _p}\text{Exp}\left[ -\overset{x,y}{\sum _{\eta }}\left(x_{i \eta}-l_{\eta}a\right)^2/\sigma ^2\right]}.
\end{equation}





and then $\tilde{U}$ becomes:

\begin{multline}
\tilde{U}= \frac{1}{2}\left(\overset{n}{\sum _i}\overset{n}{\sum _j}\overset{x,y}{\sum _{\alpha }}\overset{x,y}{\sum _{\beta }}u_{i \alpha}M_{i \alpha j \beta}u_{j \beta} + 
\right. \\ \left.
\overset{g^2}{\sum _k}\overset{g^2}{\sum _l}\overset{x,y}{\sum _{\gamma }}K_{kl}\left(\left(\overset{n}{\sum_p}u_{p \gamma}W(p,k) \right)  
\left(\overset{n}{\sum_p}u_{p \gamma}W(p,k) \right) - 
\right. \right. \\ \left.\left.
2\left(\overset{n}{\sum_p}u_{p \gamma}W(p,k) \right) 
\left(\overset{n}{\sum_p}u_{p \gamma}W(p,l) \right) +  
\right. \right.\\ \left.\left.
\left(\overset{n}{\sum_p}u_{p \gamma}W(p,l) \right) 
\left(\overset{n}{\sum_p}u_{p \gamma}W(p,l) \right)\right)\right).
\end{multline}

Without loss of generality, we reindex the summations from $p$ to $i$ and $j$
\begin{multline}
\tilde{U}= \frac{1}{2}\left(\overset{n}{\sum _i}\overset{n}{\sum _j}\overset{x,y}{\sum _{\alpha }}\overset{x,y}{\sum _{\beta }}u_{i \alpha}M_{i \alpha j \beta}u_{j \beta}+ \right.\\ \left.\overset{g^2}{\sum _k}\overset{g^2}{\sum _l}\overset{x,y}{\sum _{\gamma }}K_{kl}\left(\left(\overset{n}{\sum_i}u_{i \gamma}W(i,k) \right)\left(\overset{n}{\sum_j}u_{j \gamma}W(j,k) \right)- 
\right. \right.\\ \left.\left. 
2 \left(\overset{n}{\sum_i}u_{i \gamma}W(i,k) \right)\left(\overset{n}{\sum_j}u_{j \gamma}W(j,l) \right)+ 
\right. \right.\\ \left.\left.
\left(\overset{n}{\sum_i}u_{i \gamma}W(i,l) \right)\left(\overset{n}{\sum_j}u_{j \gamma}W(j,l) \right)\right)\right).
\end{multline}

As each term has a sum over $i$ and $j$, these can be grouped




\begin{multline}\tilde{U}= \frac{1}{2}\overset{n}{\sum _i}\overset{n}{\sum _j}\Bigg(\overset{x,y}{\sum _{\alpha }}\overset{x,y}{\sum _{\beta }}u_{i \alpha}M_{i \alpha j \beta}u_{j \beta}+ \\
\overset{g^2}{\sum _k}\overset{g^2}{\sum _l}\overset{x,y}{\sum _{\gamma }}K_{kl}u_{i \gamma}u_{j \gamma}\Big(W(i,k) W(j,k) - \\
2W(i,k) W(j,l)+
W(i,l) W(j,l) \Big)\Bigg).
\end{multline}

We reindex again:

\begin{equation}
\sum _{\gamma }u_{i \gamma}u_{j \gamma}=\sum _{\alpha }u_{i \alpha}u_{j \alpha},
\end{equation}
\begin{equation}
\sum _{\alpha }u_{i \alpha}u_{j \alpha}=\sum _{\alpha }\sum _{\beta }u_{i \alpha}u_{j \beta}\delta _{\alpha \beta},
\end{equation}

by definition of  Kronecker delta.
Grouping the dimension sums and gathering terms we find



\begin{multline}\tilde{U}= \frac{1}{2}\overset{n}{\sum _i}\overset{n}{\sum _j}\overset{x,y}{\sum _{\alpha }}\overset{x,y}{\sum _{\beta }}\Bigg(u_{i \alpha}\bigg(M_{i \alpha j \beta}+ \\
\overset{g^2}{\delta _{\alpha \beta }\sum _k}\overset{g^2}{\sum _l}K_{kl}\Big(W(i,k) *W(j,k) - \\
2W(i,k)* W(j,l) +W(i,l) *W(j,l) \Big)\bigg)u_{j \beta}\Bigg)\end{multline},

resulting in the final definition of $M^{\dagger }$,

\begin{multline}
M^{\dagger }=\overset{g^2}{\sum _k}\overset{g^2}{\sum _l}K_{kl}\delta _{\alpha \beta } \big(W(i,k)W(j,k)\\-2W(i,k)W(j,l)+W(i,l)W(j,l) \big).
\label{M_dagger_aug}
\end{multline}


We can analytically determine the energy increase for plane waves by assuming a continuous field of particles, which allows us to go from a summation to an integral over particle positions.
We let the plane wave be defined as
\begin{equation}
u_x=A \sin(2 \pi k x /L +\phi),
u_y=0.
\end{equation}
Then the continuous form is
\begin{multline}
U^\dagger = \sum_{k_x=1}^{Gt} \sum_{l_x=1}^{Gt} \sum_{k_y=1}^{Gt} \sum_{l_y=1}^{Gt} \Bigg(  \delta (l_x,k_x \pm 1) \delta (l_y,k_y \pm 1) K    \\ 
 \int_{-\infty}^{\infty} \int_{-\infty}^{\infty} \left( dx dy \frac{A}{\pi \sigma^2} \sin(2 \pi k x /L +\phi)    \right. \\ \left.
 \left( e^{-\frac{(x-a k_x)^2+(y-a k_y)^2}{\sigma ^2}}-e^{-\frac{(x-a l_x)^2+(y-a l_y)^2}{\sigma ^2}} \right) \right) \Bigg).
\end{multline}

In order to deal with boundary conditions, we take the limit as $Gt \to \infty$. 
we also set $\sigma=a$, and $\lambda=L/k$.  This computation gives the result

\begin{equation}
U^\dagger = 
-4 \pi ^2 A^2 K e^{-\frac{8 \pi ^4 a^2}{\lambda ^2}} \left(\cos \left(\frac{4 \pi ^2 a}{\lambda }\right)-1\right).
\label{eq_plane_anal}
\end{equation}
which is shown by the dashed lines in Fig. \ref{fig_param} in the main text

\section{\label{sec:Kscale}Appendix B}



\begin{figure}
\includegraphics[width=3.5in]{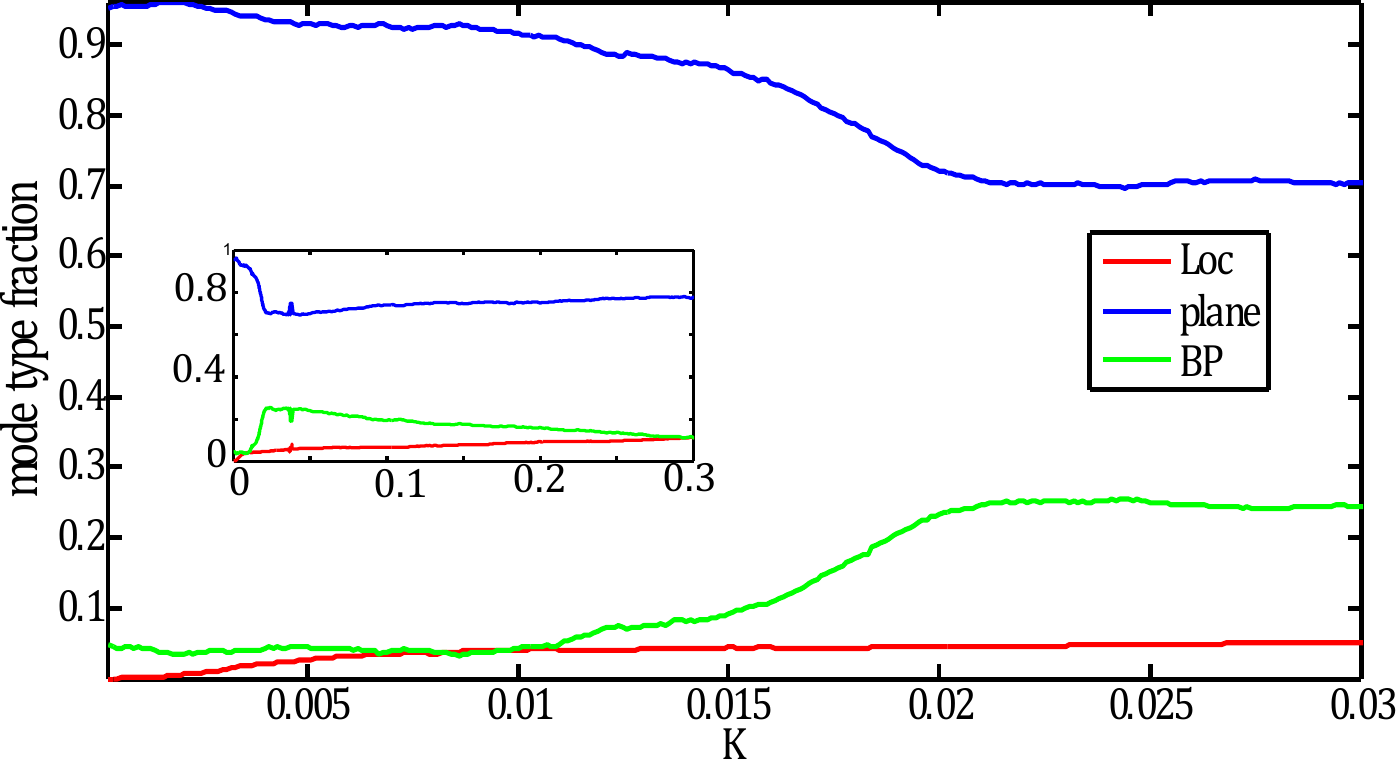}
\caption{ (color online). The 50 lowest frequency modes in 100 packings, sorted by type as in Fig.\ref{fig_2}b, plotted as a function of $K$.  There are three regions: $0<K<0.01$, where the augmentation begins to have an effect, $0.01<K<0.02$, where the augmentation begins to greatly alter the mode structure, and $0.02<K$, where the number of spurious localized modes grow linearly with $K$, as shown in the inset over a wider range of $K$.
%
}
\label{fig_param_K}
\end{figure}

To choose a value for the parameter $K$, the connection strength, we calculate the total number of localized modes of the 50 lowest frequency modes in the packing as a function of $K$, where localized modes are defined using the radius of gyration as defined in the main text. As seen in Fig \ref{fig_param_K}, the number of localized modes grows until a value of $K=0.01$, and essentially plateaus thereafter. Additionally, the number of boson peak modes is relatively constant until the same value.  Furthermore, the number of plane-wave modes begins a more sharp decline at that value.
Taken together, this suggests that a value of $K=0.01$ balances the augmented energy with the internal potential energy to shift plane waves without generating spurious localized modes.


\section{\label{sec:Energy Barriers}Appendix C}

In order to define a new state, we examine seven independent criteria: i) any change in the contact network (CR+), ii) non-rattler~\cite{rattler_note} contact changes (CR-), iii) requiring more than two particles to change contacts (C2), iv) energy differences between the original and final basins of greater than $10^{-8}$ (E-8), v) a displacement of a single particle more than two large particle diameters in a direction perpendicular to the mode (D), vi) having a significant stress drop (S), and vii) requiring that more than 2 contact changing particles must be neighbors, thereby rearranging as a unit, (C2N).  

\begin{figure}
\includegraphics[width=3.5in]{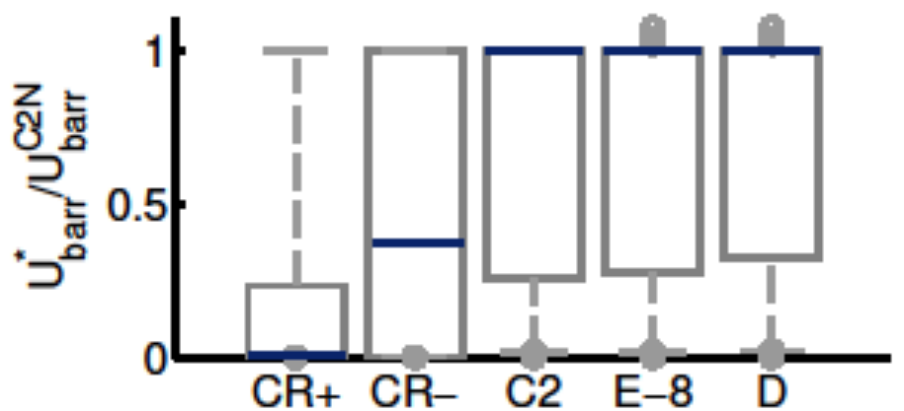}
\caption{ (color online). 
Ratio of energy barriers $\Delta U/\Delta U^{C2N}$ calculated using different definitions for what constitutes a particle rearrangement, as described in the main text. Box and whiskers contains 50 \% and 92 \% of the data points, respectively, blue bars denote the median, and outliers are circles. 
}
\label{fig_comp_barr}
\end{figure}

As shown in Fig. \ref{fig_comp_barr}, for each standard mode and each of the first five definitions, we measure the ratio between the energy barrier calculated using that definition and C2N  ($\Delta U/\Delta U^{C2N}$).Several of these criteria, such as contact changes that include or exclude rattlers (CR+, CR-), generate energy barriers that are significantly lower than the other criteria and different from each other. We note that these definitions have been used for studies of energy barriers in the past~\cite{Xu2010}.  This result suggests that these criteria generate a lot of ``false positives'' -- they identify changes to the network that do not correspond to a particle rearrangement.  In contrast, the four other methods (C2, E-8, D, S) generate distributions of energy barriers with the same median as C2N,  indicating that these criteria are very similar and likely identify particle rearrangements associated with saddle points and plasticity.

\section{\label{sec:Energy Barriers}Appendix D}

As discussed in the main text, we find that the augmented modes are just as predictive as standard modes if we use the soft spot algorithm to correlate localized excitations with plasticity, and we can use fewer modes (15 augmented modes compared to 25 standard modes).

As a third measure of whether localized excitations predict plastic events, we compare the distance between the center of mass of the localized excitation and the center of mass of the rearrangement.

Of course, we expect that this distance will scale with the number density of candidate defects, so we first calculate the probability density $P$ for the minimum distance $r$ between the origin and $n$ points randomly distributed in 2D space:
\begin{multline}
P(r,n)=2 \pi  n r \Theta \left(\frac{1}{2}-r\right) \left(1-\pi  r^2\right)^{n-1}-n \Theta \left(r-\frac{1}{2}\right) \\
* \left(-\frac{4 r}{\sqrt{4 r^2-1}}+\frac{4}{\sqrt{4-\frac{1}{r^2}}}+2 r \left(\pi -4 \csc ^{-1}(2 r)\right)\right) \\
* \left(-\sqrt{4 r^2-1}+r^2 \left(\pi -4 \csc ^{-1}(2 r)\right)+1\right)^{n-1},
\label{rand_dist_eq}
\end{multline}
   where $\Theta$ is the Heavyside function. The expected minimum distance for $n$ randomly distributed points is then $d^n_{rand}$ = $\int r P(r,n) dr.$
   
  Given the location of the center of a rearrangement and a list of $n$ locations corresponding to centers of localized excitations, we compute the minimum distance between the rearrangement and any localized excitation $d^n$.  We then normalize $d^n$ by the value of a random distribution, $d^n_{rand}$, and if our rearrangements are predictive then $d \equiv d^n/d^n_{rand} < 1$, and there is no bias as a function of the number or size of the localized regions.

  We compare the distributions of $d$ for several different definitions of localized excitations.  We first assume every localized augmented mode corresponds to a defect, and compare $d$ just before (at a strain $10^{-6}$ below the critical strain) the event (denoted 'near'), as well as immediately after the previous rearrangement (far).  We repeat this procedure for the standard modes and soft spots generated by the method of Manning and Liu~\cite{Manning2007}.

\begin{figure}
\includegraphics[width=3.5in]{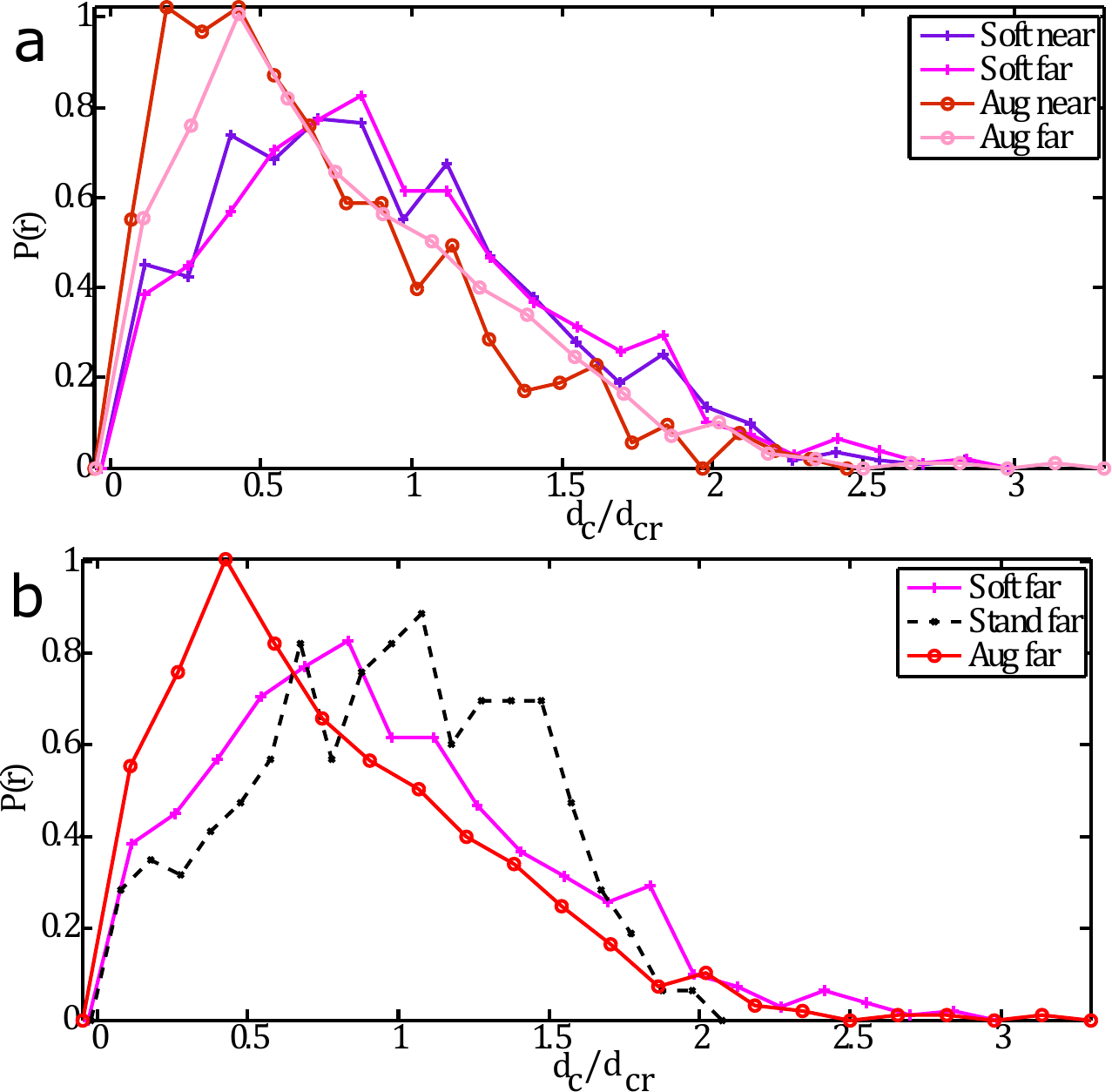}
\caption{ (color online).  Comparison of the separation distances between the center of a mode and the center of a rearrangement. We compare standard modes, augmented modes, and soft spots, near and far from the rearrangement, as well as a comparison to a random distribution of the same number as described in Appendix C.  Each line is scaled by the expected value of a random distribution of the same number of candidates.
}
\label{fig_seps}
\end{figure}

As shown in Fig. ~\ref{fig_seps}, using this strict metric, the augmented modes display significant improvement over a random distribution with the same number of candidates. The augmented modes are also closer in distance to the rearrangement than soft spots once controlled for the number of candidates. 

\begin{figure}
\includegraphics[width=3.5in]{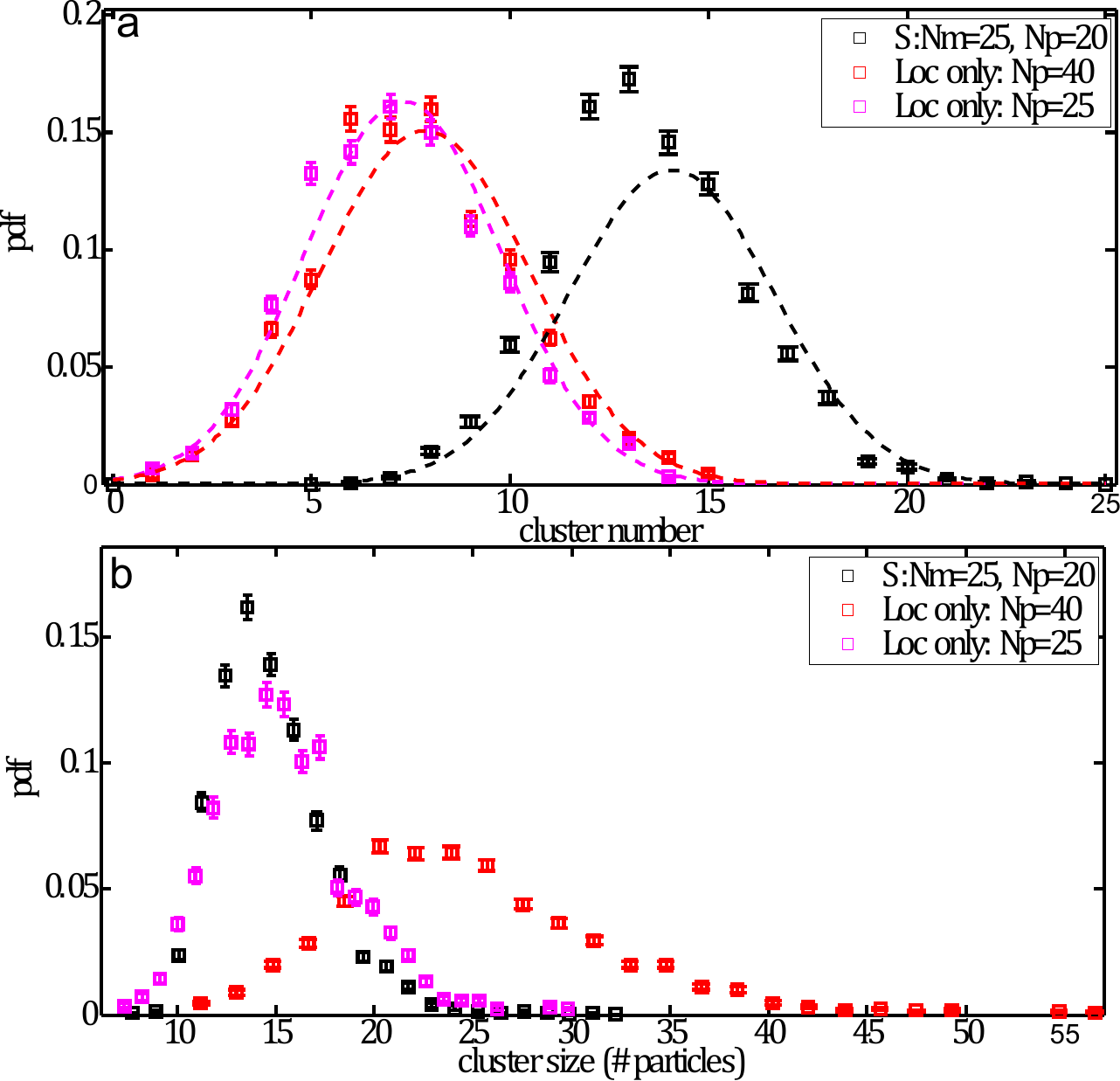}
\caption{ (color online). a)  The number of clusters for the  standard soft spots algorithm (black), as well as when using only the localized augmented modes ($R_G<16$), with $N_p=25$ (magenta) and $N_p=40$ (red).  Using only the localized modes gives fewer spots. Dashed lines are Gaussian fits.
b) The average number of particles per soft spot. Colors are same as in a) .
}
\label{fig_cluster_num_size}
\end{figure}

As discussed in the main text, we also want to understand how the size of localized augmented excitations compares to the size of soft spots.  To this end, we use the published soft spots algorithm to identify the optimal number of modes ($N_m=25$) and number of particles $(N_p=20$.  A histogram for the number of spots and their size are shown by the black data points in Fig. \ref{fig_cluster_num_size} a and b, respectively.  Next, we only study soft spots generated from localized augmented modes, and since there are only ~7 such modes we expect to find approximately that number of soft spots, which is the case, as shown by the red and magenta data points in Fig \ref{fig_cluster_num_size}a.  The correlation with rearrangements is largely insensitive to $N_p$ for values between 20 and 30, but $N_p$ does affect soft spot size, as shown in Fig. \ref{fig_cluster_num_size}b.

%



%


\bibliography{Mendeley_edited}

\begin{thebibliography}{52}
\expandafter\ifx\csname natexlab\endcsname\relax\def\natexlab#1{#1}\fi
\expandafter\ifx\csname bibnamefont\endcsname\relax
  \def\bibnamefont#1{#1}\fi
\expandafter\ifx\csname bibfnamefont\endcsname\relax
  \def\bibfnamefont#1{#1}\fi
\expandafter\ifx\csname citenamefont\endcsname\relax
  \def\citenamefont#1{#1}\fi
\expandafter\ifx\csname url\endcsname\relax
  \def\url#1{\texttt{#1}}\fi
\expandafter\ifx\csname urlprefix\endcsname\relax\def\urlprefix{URL }\fi
\providecommand{\bibinfo}[2]{#2}
\providecommand{\eprint}[2][]{\url{#2}}

\bibitem[{\citenamefont{Taylor}(1934)}]{Taylor1934}
\bibinfo{author}{\bibfnamefont{G.~I.} \bibnamefont{Taylor}},
  \bibinfo{journal}{Proceedings of the Royal Society A: Mathematical, Physical
  and Engineering Sciences} \textbf{\bibinfo{volume}{145}},
  \bibinfo{pages}{362} (\bibinfo{year}{1934}), ISSN \bibinfo{issn}{1364-5021},
  \urlprefix\url{http://rspa.royalsocietypublishing.org/cgi/doi/10.1098/rspa.1934.0106}.

\bibitem[{\citenamefont{Slotterback et~al.}(2012)\citenamefont{Slotterback,
  Mailman, Ronaszegi, Van~Hecke, Girvan, and Losert}}]{Slotterback2012}
\bibinfo{author}{\bibfnamefont{S.}~\bibnamefont{Slotterback}},
  \bibinfo{author}{\bibfnamefont{M.}~\bibnamefont{Mailman}},
  \bibinfo{author}{\bibfnamefont{K.}~\bibnamefont{Ronaszegi}},
  \bibinfo{author}{\bibfnamefont{M.}~\bibnamefont{Van~Hecke}},
  \bibinfo{author}{\bibfnamefont{M.}~\bibnamefont{Girvan}}, \bibnamefont{and}
  \bibinfo{author}{\bibfnamefont{W.}~\bibnamefont{Losert}},
  \bibinfo{journal}{Physical Review E - Statistical, Nonlinear, and Soft Matter
  Physics} \textbf{\bibinfo{volume}{85}}, \bibinfo{pages}{021309}
  (\bibinfo{year}{2012}), ISSN \bibinfo{issn}{15393755},
  \urlprefix\url{http://link.aps.org/doi/10.1103/PhysRevE.85.021309}.

\bibitem[{\citenamefont{Henann and Kamrin}(2014)}]{Henann2014}
\bibinfo{author}{\bibfnamefont{D.~L.} \bibnamefont{Henann}} \bibnamefont{and}
  \bibinfo{author}{\bibfnamefont{K.}~\bibnamefont{Kamrin}},
  \bibinfo{journal}{Physical Review Letters} \textbf{\bibinfo{volume}{113}},
  \bibinfo{pages}{178001} (\bibinfo{year}{2014}), ISSN
  \bibinfo{issn}{0031-9007},
  \urlprefix\url{http://link.aps.org/doi/10.1103/PhysRevLett.113.178001}.

\bibitem[{\citenamefont{Coulais et~al.}(2014)\citenamefont{Coulais, Seguin, and
  Dauchot}}]{Coulais2014}
\bibinfo{author}{\bibfnamefont{C.}~\bibnamefont{Coulais}},
  \bibinfo{author}{\bibfnamefont{a.}~\bibnamefont{Seguin}}, \bibnamefont{and}
  \bibinfo{author}{\bibfnamefont{O.}~\bibnamefont{Dauchot}},
  \bibinfo{journal}{Physical Review Letters} \textbf{\bibinfo{volume}{113}},
  \bibinfo{pages}{1} (\bibinfo{year}{2014}), ISSN \bibinfo{issn}{10797114},
  \urlprefix\url{http://link.aps.org/doi/10.1103/PhysRevLett.113.198001}.

\bibitem[{\citenamefont{Morgan and Boettcher}(1999)}]{Morgan1999}
\bibinfo{author}{\bibfnamefont{J.~K.} \bibnamefont{Morgan}} \bibnamefont{and}
  \bibinfo{author}{\bibfnamefont{M.~S.} \bibnamefont{Boettcher}},
  \bibinfo{journal}{Journal of Geophysical Research}
  \textbf{\bibinfo{volume}{104}}, \bibinfo{pages}{2703} (\bibinfo{year}{1999}),
  ISSN \bibinfo{issn}{0148-0227}.

\bibitem[{\citenamefont{Su et~al.}(2012)\citenamefont{Su, Liu, and
  Lin}}]{Su2012}
\bibinfo{author}{\bibfnamefont{Y.~S.} \bibnamefont{Su}},
  \bibinfo{author}{\bibfnamefont{Y.~H.} \bibnamefont{Liu}}, \bibnamefont{and}
  \bibinfo{author}{\bibfnamefont{I.}~\bibnamefont{Lin}},
  \bibinfo{journal}{Physical Review Letters} \textbf{\bibinfo{volume}{109}},
  \bibinfo{pages}{195002} (\bibinfo{year}{2012}), ISSN
  \bibinfo{issn}{00319007},
  \urlprefix\url{http://link.aps.org/doi/10.1103/PhysRevLett.109.195002}.

\bibitem[{\citenamefont{Lerner et~al.}(2013)\citenamefont{Lerner, D{\"{u}}ring,
  and Wyart}}]{Lerner2013}
\bibinfo{author}{\bibfnamefont{E.}~\bibnamefont{Lerner}},
  \bibinfo{author}{\bibfnamefont{G.}~\bibnamefont{D{\"{u}}ring}},
  \bibnamefont{and} \bibinfo{author}{\bibfnamefont{M.}~\bibnamefont{Wyart}},
  \bibinfo{journal}{Soft Matter} \textbf{\bibinfo{volume}{9}},
  \bibinfo{pages}{8252} (\bibinfo{year}{2013}), ISSN \bibinfo{issn}{1744-683X},
  \urlprefix\url{http://xlink.rsc.org/?DOI=c3sm50515d}.

\bibitem[{\citenamefont{Schall et~al.}(2007)\citenamefont{Schall, Weitz, and
  Spaepen}}]{Schall2007}
\bibinfo{author}{\bibfnamefont{P.}~\bibnamefont{Schall}},
  \bibinfo{author}{\bibfnamefont{D.~A.} \bibnamefont{Weitz}}, \bibnamefont{and}
  \bibinfo{author}{\bibfnamefont{F.}~\bibnamefont{Spaepen}},
  \bibinfo{journal}{Science} \textbf{\bibinfo{volume}{318}},
  \bibinfo{pages}{1895} (\bibinfo{year}{2007}), ISSN \bibinfo{issn}{1095-9203},
  \urlprefix\url{http://www.ncbi.nlm.nih.gov/pubmed/18096800}.

\bibitem[{\citenamefont{Falk and Langer}(1997)}]{Falk1998}
\bibinfo{author}{\bibfnamefont{M.~L.} \bibnamefont{Falk}} \bibnamefont{and}
  \bibinfo{author}{\bibfnamefont{J.~S.} \bibnamefont{Langer}},
  \bibinfo{journal}{Physical Review E} \textbf{\bibinfo{volume}{57}},
  \bibinfo{pages}{16} (\bibinfo{year}{1997}), ISSN \bibinfo{issn}{1063-651X},
  \urlprefix\url{http://arxiv.org/abs/cond-mat/9712114}.

\bibitem[{\citenamefont{Manning et~al.}(2007)\citenamefont{Manning, Langer, and
  Carlson}}]{Manning2007}
\bibinfo{author}{\bibfnamefont{M.~L.} \bibnamefont{Manning}},
  \bibinfo{author}{\bibfnamefont{J.~S.} \bibnamefont{Langer}},
  \bibnamefont{and} \bibinfo{author}{\bibfnamefont{J.~M.}
  \bibnamefont{Carlson}}, \bibinfo{journal}{Physical Review E - Statistical,
  Nonlinear, and Soft Matter Physics} \textbf{\bibinfo{volume}{76}},
  \bibinfo{pages}{056106} (\bibinfo{year}{2007}), ISSN
  \bibinfo{issn}{15393755},
  \urlprefix\url{http://pre.aps.org/abstract/PRE/v76/i5/e056106}.

\bibitem[{\citenamefont{Martens et~al.}(2012)\citenamefont{Martens, Bocquet,
  and Barrat}}]{Martens2012}
\bibinfo{author}{\bibfnamefont{K.}~\bibnamefont{Martens}},
  \bibinfo{author}{\bibfnamefont{L.}~\bibnamefont{Bocquet}}, \bibnamefont{and}
  \bibinfo{author}{\bibfnamefont{J.-L.} \bibnamefont{Barrat}},
  \bibinfo{journal}{Soft Matter} \textbf{\bibinfo{volume}{8}},
  \bibinfo{pages}{4197} (\bibinfo{year}{2012}), ISSN \bibinfo{issn}{1744-683X}.

\bibitem[{\citenamefont{Salerno et~al.}(2012)\citenamefont{Salerno, Maloney,
  and Robbins}}]{Salerno2012}
\bibinfo{author}{\bibfnamefont{K.~M.} \bibnamefont{Salerno}},
  \bibinfo{author}{\bibfnamefont{C.~E.} \bibnamefont{Maloney}},
  \bibnamefont{and} \bibinfo{author}{\bibfnamefont{M.~O.}
  \bibnamefont{Robbins}}, \bibinfo{journal}{Physical Review Letters}
  \textbf{\bibinfo{volume}{109}}, \bibinfo{pages}{105703}
  (\bibinfo{year}{2012}), ISSN \bibinfo{issn}{00319007},
  \urlprefix\url{http://link.aps.org/doi/10.1103/PhysRevLett.109.105703}.

\bibitem[{\citenamefont{Keim and Arratia}(2013)}]{Keim2013}
\bibinfo{author}{\bibfnamefont{N.~C.} \bibnamefont{Keim}} \bibnamefont{and}
  \bibinfo{author}{\bibfnamefont{P.~E.} \bibnamefont{Arratia}},
  \bibinfo{journal}{Soft Matter} \textbf{\bibinfo{volume}{9}},
  \bibinfo{pages}{6222} (\bibinfo{year}{2013}), ISSN \bibinfo{issn}{1744-683X},
  \urlprefix\url{http://xlink.rsc.org/?DOI=c3sm51014j\npapers3://publication/doi/10.1039/c3sm51014j
  http://www.scopus.com/inward/record.url?eid=2-s2.0-84881055679&partnerID=tZOtx3y1}.

\bibitem[{\citenamefont{Fiocco et~al.}(2014)\citenamefont{Fiocco, Foffi, and
  Sastry}}]{Fiocco2014}
\bibinfo{author}{\bibfnamefont{D.}~\bibnamefont{Fiocco}},
  \bibinfo{author}{\bibfnamefont{G.}~\bibnamefont{Foffi}}, \bibnamefont{and}
  \bibinfo{author}{\bibfnamefont{S.}~\bibnamefont{Sastry}},
  \bibinfo{journal}{Physical Review Letters} \textbf{\bibinfo{volume}{112}},
  \bibinfo{pages}{025702} (\bibinfo{year}{2014}), ISSN
  \bibinfo{issn}{00319007},
  \urlprefix\url{http://link.aps.org/doi/10.1103/PhysRevLett.112.025702}.

\bibitem[{\citenamefont{Falk and Langer}(2010)}]{Falk2011}
\bibinfo{author}{\bibfnamefont{M.~L.} \bibnamefont{Falk}} \bibnamefont{and}
  \bibinfo{author}{\bibfnamefont{J.~S.} \bibnamefont{Langer}},
  \bibinfo{journal}{Annual Review of Condensed Matter Physics}
  \textbf{\bibinfo{volume}{2}}, \bibinfo{pages}{28} (\bibinfo{year}{2010}),
  ISSN \bibinfo{issn}{1947-5454},
  \urlprefix\url{http://arxiv.org/abs/1004.4684}.

\bibitem[{\citenamefont{Sollich et~al.}(1996)\citenamefont{Sollich, Lequeux,
  Hebraud, and Cates}}]{Sollich1997}
\bibinfo{author}{\bibfnamefont{P.}~\bibnamefont{Sollich}},
  \bibinfo{author}{\bibfnamefont{F.}~\bibnamefont{Lequeux}},
  \bibinfo{author}{\bibfnamefont{P.}~\bibnamefont{Hebraud}}, \bibnamefont{and}
  \bibinfo{author}{\bibfnamefont{M.~E.} \bibnamefont{Cates}},
  \bibinfo{journal}{Physical Review Letters} \textbf{\bibinfo{volume}{78}},
  \bibinfo{pages}{4} (\bibinfo{year}{1996}), ISSN \bibinfo{issn}{0295-5075},
  \urlprefix\url{http://arxiv.org/abs/cond-mat/9611228}.

\bibitem[{\citenamefont{Widmer-Cooper et~al.}(2009)\citenamefont{Widmer-Cooper,
  Perry, Harrowell, and Reichman}}]{Widmer-Cooper2008}
\bibinfo{author}{\bibfnamefont{A.}~\bibnamefont{Widmer-Cooper}},
  \bibinfo{author}{\bibfnamefont{H.}~\bibnamefont{Perry}},
  \bibinfo{author}{\bibfnamefont{P.}~\bibnamefont{Harrowell}},
  \bibnamefont{and} \bibinfo{author}{\bibfnamefont{D.~R.}
  \bibnamefont{Reichman}}, \bibinfo{journal}{Nature Physics}
  \textbf{\bibinfo{volume}{4}}, \bibinfo{pages}{711} (\bibinfo{year}{2009}),
  ISSN \bibinfo{issn}{1745-2473},
  \urlprefix\url{http://arxiv.org/abs/0901.3547}.

\bibitem[{\citenamefont{Tanguy et~al.}(2010)\citenamefont{Tanguy, Mantisi, and
  Tsamados}}]{Tanguy2010}
\bibinfo{author}{\bibfnamefont{a.}~\bibnamefont{Tanguy}},
  \bibinfo{author}{\bibfnamefont{B.}~\bibnamefont{Mantisi}}, \bibnamefont{and}
  \bibinfo{author}{\bibfnamefont{M.}~\bibnamefont{Tsamados}},
  \bibinfo{journal}{EPL (Europhysics Letters)} \textbf{\bibinfo{volume}{90}},
  \bibinfo{pages}{16004} (\bibinfo{year}{2010}), ISSN
  \bibinfo{issn}{0295-5075},
  \urlprefix\url{http://stacks.iop.org/0295-5075/90/i=1/a=16004}.

\bibitem[{\citenamefont{Manning and Liu}(2011)}]{Manning2011}
\bibinfo{author}{\bibfnamefont{M.~L.} \bibnamefont{Manning}} \bibnamefont{and}
  \bibinfo{author}{\bibfnamefont{A.~J.} \bibnamefont{Liu}},
  \bibinfo{journal}{Physical Review Letters} \textbf{\bibinfo{volume}{107}}
  (\bibinfo{year}{2011}), ISSN \bibinfo{issn}{00319007},
  \urlprefix\url{http://prl.aps.org/abstract/PRL/v107/i10/e108302
  http://journals.aps.org/prl/abstract/10.1103/PhysRevLett.107.108302}.

\bibitem[{\citenamefont{Schoenholz et~al.}(2014)\citenamefont{Schoenholz, Liu,
  Riggleman, and Rottler}}]{Riggleman2014}
\bibinfo{author}{\bibfnamefont{S.~S.} \bibnamefont{Schoenholz}},
  \bibinfo{author}{\bibfnamefont{A.~J.} \bibnamefont{Liu}},
  \bibinfo{author}{\bibfnamefont{R.~a.} \bibnamefont{Riggleman}},
  \bibnamefont{and} \bibinfo{author}{\bibfnamefont{J.}~\bibnamefont{Rottler}},
  \bibinfo{journal}{Physical Review X} \textbf{\bibinfo{volume}{4}},
  \bibinfo{pages}{031014} (\bibinfo{year}{2014}), ISSN
  \bibinfo{issn}{21603308},
  \urlprefix\url{http://link.aps.org/doi/10.1103/PhysRevX.4.031014
  http://arxiv.org/abs/1404.1403}.

\bibitem[{\citenamefont{Sussman et~al.}(2015)\citenamefont{Sussman, Goodrich,
  Liu, and Nagel}}]{Sussman2014}
\bibinfo{author}{\bibfnamefont{D.~M.} \bibnamefont{Sussman}},
  \bibinfo{author}{\bibfnamefont{C.~P.} \bibnamefont{Goodrich}},
  \bibinfo{author}{\bibfnamefont{A.~J.} \bibnamefont{Liu}}, \bibnamefont{and}
  \bibinfo{author}{\bibfnamefont{S.~R.} \bibnamefont{Nagel}},
  \bibinfo{journal}{Soft Matter} \textbf{\bibinfo{volume}{11}},
  \bibinfo{pages}{2745} (\bibinfo{year}{2015}), ISSN \bibinfo{issn}{1744-683X},
  \urlprefix\url{http://arxiv.org/abs/1412.8755
  http://xlink.rsc.org/?DOI=C4SM02905D}.

\bibitem[{\citenamefont{Rottler et~al.}(2014)\citenamefont{Rottler, Schoenholz,
  and Liu}}]{Schoenholz2014}
\bibinfo{author}{\bibfnamefont{J.}~\bibnamefont{Rottler}},
  \bibinfo{author}{\bibfnamefont{S.~S.} \bibnamefont{Schoenholz}},
  \bibnamefont{and} \bibinfo{author}{\bibfnamefont{A.~J.} \bibnamefont{Liu}},
  \bibinfo{journal}{Physical Review E - Statistical, Nonlinear, and Soft Matter
  Physics} \textbf{\bibinfo{volume}{89}}, \bibinfo{pages}{2}
  (\bibinfo{year}{2014}), ISSN \bibinfo{issn}{15502376},
  \urlprefix\url{http://arxiv.org/abs/1403.0922
  http://journals.aps.org/pre/abstract/10.1103/PhysRevE.89.042304}.

\bibitem[{\citenamefont{Cubuk et~al.}(2015)\citenamefont{Cubuk, Schoenholz,
  Rieser, Malone, Rottler, Durian, Kaxiras, and Liu}}]{Cubuk2015}
\bibinfo{author}{\bibfnamefont{E.~D.} \bibnamefont{Cubuk}},
  \bibinfo{author}{\bibfnamefont{S.~S.} \bibnamefont{Schoenholz}},
  \bibinfo{author}{\bibfnamefont{J.~M.} \bibnamefont{Rieser}},
  \bibinfo{author}{\bibfnamefont{B.~D.} \bibnamefont{Malone}},
  \bibinfo{author}{\bibfnamefont{J.}~\bibnamefont{Rottler}},
  \bibinfo{author}{\bibfnamefont{D.~J.} \bibnamefont{Durian}},
  \bibinfo{author}{\bibfnamefont{E.}~\bibnamefont{Kaxiras}}, \bibnamefont{and}
  \bibinfo{author}{\bibfnamefont{A.~J.} \bibnamefont{Liu}},
  \bibinfo{journal}{Physical Review Letters} \textbf{\bibinfo{volume}{114}},
  \bibinfo{pages}{108001} (\bibinfo{year}{2015}), ISSN
  \bibinfo{issn}{10797114},
  \urlprefix\url{http://link.aps.org/doi/10.1103/PhysRevLett.114.108001}.

\bibitem[{\citenamefont{Gartner and Lerner}(2016{\natexlab{a}})}]{Gartner2016}
\bibinfo{author}{\bibfnamefont{L.}~\bibnamefont{Gartner}} \bibnamefont{and}
  \bibinfo{author}{\bibfnamefont{E.}~\bibnamefont{Lerner}},
  \bibinfo{journal}{Physical Review E} \textbf{\bibinfo{volume}{93}},
  \bibinfo{pages}{8} (\bibinfo{year}{2016}{\natexlab{a}}), ISSN
  \bibinfo{issn}{2470-0045},
  \urlprefix\url{http://journals.aps.org/pre/abstract/10.1103/PhysRevE.93.011001}.

\bibitem[{\citenamefont{Patinet et~al.}(2016)\citenamefont{Patinet,
  Vandembroucq, and Falk}}]{Patinet2016}
\bibinfo{author}{\bibfnamefont{S.}~\bibnamefont{Patinet}},
  \bibinfo{author}{\bibfnamefont{D.}~\bibnamefont{Vandembroucq}},
  \bibnamefont{and} \bibinfo{author}{\bibfnamefont{M.~L.} \bibnamefont{Falk}},
  \bibinfo{journal}{Physical Review Letters} \textbf{\bibinfo{volume}{117}},
  \bibinfo{pages}{045501} (\bibinfo{year}{2016}), ISSN
  \bibinfo{issn}{0031-9007},
  \urlprefix\url{http://link.aps.org/doi/10.1103/PhysRevLett.117.045501}.

\bibitem[{\citenamefont{Xu et~al.}(2009)\citenamefont{Xu, Vitelli, Liu, and
  Nagel}}]{Xu2010}
\bibinfo{author}{\bibfnamefont{N.}~\bibnamefont{Xu}},
  \bibinfo{author}{\bibfnamefont{V.}~\bibnamefont{Vitelli}},
  \bibinfo{author}{\bibfnamefont{A.~J.} \bibnamefont{Liu}}, \bibnamefont{and}
  \bibinfo{author}{\bibfnamefont{S.~R.} \bibnamefont{Nagel}},
  \bibinfo{journal}{EPL (Europhysics Letters)} \textbf{\bibinfo{volume}{90}},
  \bibinfo{pages}{6} (\bibinfo{year}{2009}), ISSN \bibinfo{issn}{0295-5075},
  \urlprefix\url{http://arxiv.org/abs/0909.3701}.

\bibitem[{\citenamefont{Morse et~al.}(2017)\citenamefont{Morse, Wijtmans, Deen,
  Hecke, and Manning}}]{Morse2017}
\bibinfo{author}{\bibfnamefont{P.}~\bibnamefont{Morse}},
  \bibinfo{author}{\bibfnamefont{S.}~\bibnamefont{Wijtmans}},
  \bibinfo{author}{\bibfnamefont{M.~V.} \bibnamefont{Deen}},
  \bibinfo{author}{\bibfnamefont{M.~V.} \bibnamefont{Hecke}}, \bibnamefont{and}
  \bibinfo{author}{\bibfnamefont{M.~L.} \bibnamefont{Manning}},
  \bibinfo{journal}{In preperation} pp. \bibinfo{pages}{2--7}
  (\bibinfo{year}{2017}).

\bibitem[{\citenamefont{Zheng et~al.}(2016)\citenamefont{Zheng, Liu, and
  Xu}}]{Zheng2016}
\bibinfo{author}{\bibfnamefont{W.}~\bibnamefont{Zheng}},
  \bibinfo{author}{\bibfnamefont{H.}~\bibnamefont{Liu}}, \bibnamefont{and}
  \bibinfo{author}{\bibfnamefont{N.}~\bibnamefont{Xu}},
  \bibinfo{journal}{Physical Review E} \textbf{\bibinfo{volume}{94}},
  \bibinfo{pages}{062608} (\bibinfo{year}{2016}), ISSN
  \bibinfo{issn}{2470-0045},
  \urlprefix\url{http://link.aps.org/doi/10.1103/PhysRevE.94.062608}.

\bibitem[{\citenamefont{O’Hern et~al.}(2003)\citenamefont{O’Hern, Silbert,
  and Nagel}}]{OHern2003}
\bibinfo{author}{\bibfnamefont{C.~S.} \bibnamefont{O’Hern}},
  \bibinfo{author}{\bibfnamefont{L.~E.} \bibnamefont{Silbert}},
  \bibnamefont{and} \bibinfo{author}{\bibfnamefont{S.~R.} \bibnamefont{Nagel}},
  \bibinfo{journal}{Physical Review E} \textbf{\bibinfo{volume}{68}},
  \bibinfo{pages}{11306} (\bibinfo{year}{2003}), ISSN
  \bibinfo{issn}{1539-3755}, \eprint{0304421v1},
  \urlprefix\url{http://link.aps.org/doi/10.1103/PhysRevE.68.011306}.

\bibitem[{\citenamefont{Lees and Edwards}(1972)}]{Lees1972}
\bibinfo{author}{\bibfnamefont{A.~W.} \bibnamefont{Lees}} \bibnamefont{and}
  \bibinfo{author}{\bibfnamefont{S.~F.} \bibnamefont{Edwards}},
  \bibinfo{journal}{Journal of Physics C: Solid State Physics}
  \textbf{\bibinfo{volume}{5}}, \bibinfo{pages}{1921} (\bibinfo{year}{1972}),
  ISSN \bibinfo{issn}{0022-3719},
  \urlprefix\url{http://stacks.iop.org/0022-3719/5/i=15/a=006?key=crossref.e594e5cd0b6eb0d5dc9f68b23ee836be}.

\bibitem[{ARP()}]{ARPACK}
\bibinfo{note}{Http://www.caam.rice.edu/software/ARPACK/}.

\bibitem[{\citenamefont{v.~Muench and Statz}(1966)}]{Ashcroft1976}
\bibinfo{author}{\bibfnamefont{W.}~\bibnamefont{v.~Muench}} \bibnamefont{and}
  \bibinfo{author}{\bibfnamefont{H.}~\bibnamefont{Statz}},
  \emph{\bibinfo{title}{{Solid-to-solid diffusion in the gallium arsenide
  device technology}}}, vol.~\bibinfo{volume}{9} (\bibinfo{year}{1966}), ISBN
  \bibinfo{isbn}{0030839939},
  \urlprefix\url{http://www.amazon.com/Solid-State-Physics-Neil-Ashcroft/dp/0030839939}.

\bibitem[{\citenamefont{Schreck et~al.}(2011)\citenamefont{Schreck, Bertrand,
  O’Hern, and Shattuck}}]{Schreck2011}
\bibinfo{author}{\bibfnamefont{C.~F.} \bibnamefont{Schreck}},
  \bibinfo{author}{\bibfnamefont{T.}~\bibnamefont{Bertrand}},
  \bibinfo{author}{\bibfnamefont{C.~S.} \bibnamefont{O’Hern}},
  \bibnamefont{and} \bibinfo{author}{\bibfnamefont{M.~D.}
  \bibnamefont{Shattuck}}, \bibinfo{journal}{Physical Review Letters}
  \textbf{\bibinfo{volume}{107}}, \bibinfo{pages}{078301}
  (\bibinfo{year}{2011}), ISSN \bibinfo{issn}{00319007},
  \urlprefix\url{http://link.aps.org/doi/10.1103/PhysRevLett.107.078301}.

\bibitem[{\citenamefont{Goodrich et~al.}(2014)\citenamefont{Goodrich, Liu, and
  Nagel}}]{Goodrich2014a}
\bibinfo{author}{\bibfnamefont{C.~P.} \bibnamefont{Goodrich}},
  \bibinfo{author}{\bibfnamefont{A.~J.} \bibnamefont{Liu}}, \bibnamefont{and}
  \bibinfo{author}{\bibfnamefont{S.~R.} \bibnamefont{Nagel}},
  \bibinfo{journal}{Physical Review E} \textbf{\bibinfo{volume}{90}},
  \bibinfo{pages}{022201} (\bibinfo{year}{2014}), ISSN
  \bibinfo{issn}{1539-3755},
  \urlprefix\url{http://link.aps.org/doi/10.1103/PhysRevE.90.022201}.

\bibitem[{\citenamefont{van Deen et~al.}(2014)\citenamefont{van Deen, Simon,
  Zeravcic, Dagois-Bohy, Tighe, and van Hecke}}]{VanDeen2014}
\bibinfo{author}{\bibfnamefont{M.~S.} \bibnamefont{van Deen}},
  \bibinfo{author}{\bibfnamefont{J.}~\bibnamefont{Simon}},
  \bibinfo{author}{\bibfnamefont{Z.}~\bibnamefont{Zeravcic}},
  \bibinfo{author}{\bibfnamefont{S.}~\bibnamefont{Dagois-Bohy}},
  \bibinfo{author}{\bibfnamefont{B.~P.} \bibnamefont{Tighe}}, \bibnamefont{and}
  \bibinfo{author}{\bibfnamefont{M.}~\bibnamefont{van Hecke}},
  \bibinfo{journal}{Physical Review E} \textbf{\bibinfo{volume}{90}},
  \bibinfo{pages}{020202} (\bibinfo{year}{2014}), ISSN
  \bibinfo{issn}{1539-3755},
  \urlprefix\url{http://arxiv.org/pdf/1404.3156v1.pdf
  http://link.aps.org/doi/10.1103/PhysRevE.90.020202
  http://arxiv.org/abs/1404.3156v1}.

\bibitem[{\citenamefont{Maloney and Lema{\^{i}}tre}(2006)}]{Maloney2006}
\bibinfo{author}{\bibfnamefont{C.~E.} \bibnamefont{Maloney}} \bibnamefont{and}
  \bibinfo{author}{\bibfnamefont{A.}~\bibnamefont{Lema{\^{i}}tre}},
  \bibinfo{journal}{Physical Review E - Statistical, Nonlinear, and Soft Matter
  Physics} \textbf{\bibinfo{volume}{74}}, \bibinfo{pages}{016118}
  (\bibinfo{year}{2006}), ISSN \bibinfo{issn}{15393755},
  \urlprefix\url{http://www.ncbi.nlm.nih.gov/pubmed/16907162}.

\bibitem[{\citenamefont{Donati et~al.}(1997)\citenamefont{Donati, Douglas, Kob,
  Plimpton, Poole, and Glotzer}}]{Donati1997}
\bibinfo{author}{\bibfnamefont{C.}~\bibnamefont{Donati}},
  \bibinfo{author}{\bibfnamefont{J.~F.} \bibnamefont{Douglas}},
  \bibinfo{author}{\bibfnamefont{W.}~\bibnamefont{Kob}},
  \bibinfo{author}{\bibfnamefont{S.~J.} \bibnamefont{Plimpton}},
  \bibinfo{author}{\bibfnamefont{P.~H.} \bibnamefont{Poole}}, \bibnamefont{and}
  \bibinfo{author}{\bibfnamefont{S.~C.} \bibnamefont{Glotzer}},
  \emph{\bibinfo{title}{{String-like Clusters and Cooperative Motion in a Model
  Glass-Forming Liquid}}} (\bibinfo{year}{1997}),
  \urlprefix\url{http://arxiv.org/abs/cond-mat/9706277
  http://journals.aps.org/prl/pdf/10.1103/PhysRevLett.80.2338}.

\bibitem[{\citenamefont{Manning and Liu}(2015)}]{Manning2015b}
\bibinfo{author}{\bibfnamefont{M.~L.} \bibnamefont{Manning}} \bibnamefont{and}
  \bibinfo{author}{\bibfnamefont{A.~J.} \bibnamefont{Liu}},
  \bibinfo{journal}{EPL (Europhysics Letters)} \textbf{\bibinfo{volume}{109}},
  \bibinfo{pages}{36002} (\bibinfo{year}{2015}), ISSN
  \bibinfo{issn}{0295-5075},
  \urlprefix\url{http://stacks.iop.org/0295-5075/109/i=3/a=36002?key=crossref.6feff301ac77239cfed8458d6bc8a278}.

\bibitem[{\citenamefont{Shavit et~al.}(2013)\citenamefont{Shavit, Douglas, and
  Riggleman}}]{Shavit2013}
\bibinfo{author}{\bibfnamefont{A.}~\bibnamefont{Shavit}},
  \bibinfo{author}{\bibfnamefont{J.~F.} \bibnamefont{Douglas}},
  \bibnamefont{and} \bibinfo{author}{\bibfnamefont{R.~A.}
  \bibnamefont{Riggleman}}, \bibinfo{journal}{The Journal of chemical physics}
  \textbf{\bibinfo{volume}{138}}, \bibinfo{pages}{12A528}
  (\bibinfo{year}{2013}), ISSN \bibinfo{issn}{1089-7690},
  \urlprefix\url{http://www.pubmedcentral.nih.gov/articlerender.fcgi?artid=3574088&tool=pmcentrez&rendertype=abstract}.

\bibitem[{\citenamefont{Ding et~al.}(2014)\citenamefont{Ding, Patinet, Falk,
  Cheng, and Ma}}]{Ding2014}
\bibinfo{author}{\bibfnamefont{J.}~\bibnamefont{Ding}},
  \bibinfo{author}{\bibfnamefont{S.}~\bibnamefont{Patinet}},
  \bibinfo{author}{\bibfnamefont{M.~L.} \bibnamefont{Falk}},
  \bibinfo{author}{\bibfnamefont{Y.}~\bibnamefont{Cheng}}, \bibnamefont{and}
  \bibinfo{author}{\bibfnamefont{E.}~\bibnamefont{Ma}},
  \bibinfo{journal}{Proceedings of the National Academy of Sciences}
  \textbf{\bibinfo{volume}{111}}, \bibinfo{pages}{14052}
  (\bibinfo{year}{2014}), ISSN \bibinfo{issn}{0027-8424},
  \urlprefix\url{http://www.pnas.org/cgi/doi/10.1073/pnas.1412095111}.

\bibitem[{\citenamefont{Lerner and Procaccia}(2009)}]{Lerner2009}
\bibinfo{author}{\bibfnamefont{E.}~\bibnamefont{Lerner}} \bibnamefont{and}
  \bibinfo{author}{\bibfnamefont{I.}~\bibnamefont{Procaccia}},
  \bibinfo{journal}{Physical Review E - Statistical, Nonlinear, and Soft Matter
  Physics} \textbf{\bibinfo{volume}{79}}, \bibinfo{pages}{1}
  (\bibinfo{year}{2009}), ISSN \bibinfo{issn}{15393755}.

\bibitem[{\citenamefont{Schr{\o}der et~al.}(2000)\citenamefont{Schr{\o}der,
  Sastry, Dyre, and Glotzer}}]{Schrder2000}
\bibinfo{author}{\bibfnamefont{T.~B.} \bibnamefont{Schr{\o}der}},
  \bibinfo{author}{\bibfnamefont{S.}~\bibnamefont{Sastry}},
  \bibinfo{author}{\bibfnamefont{J.~C.} \bibnamefont{Dyre}}, \bibnamefont{and}
  \bibinfo{author}{\bibfnamefont{S.~C.} \bibnamefont{Glotzer}},
  \bibinfo{journal}{Journal of Chemical Physics}
  \textbf{\bibinfo{volume}{112}}, \bibinfo{pages}{9834} (\bibinfo{year}{2000}),
  ISSN \bibinfo{issn}{00219606}.

\bibitem[{\citenamefont{Donati et~al.}(1998)\citenamefont{Donati, Douglas, Kob,
  Plimpton, Poole, and Glotzer}}]{Donati1998}
\bibinfo{author}{\bibfnamefont{C.}~\bibnamefont{Donati}},
  \bibinfo{author}{\bibfnamefont{J.}~\bibnamefont{Douglas}},
  \bibinfo{author}{\bibfnamefont{W.}~\bibnamefont{Kob}},
  \bibinfo{author}{\bibfnamefont{S.}~\bibnamefont{Plimpton}},
  \bibinfo{author}{\bibfnamefont{P.}~\bibnamefont{Poole}}, \bibnamefont{and}
  \bibinfo{author}{\bibfnamefont{S.}~\bibnamefont{Glotzer}},
  \emph{\bibinfo{title}{{Stringlike Cooperative Motion in a Supercooled
  Liquid}}} (\bibinfo{year}{1998}),
  \urlprefix\url{http://journals.aps.org/prl/pdf/10.1103/PhysRevLett.80.2338}.

\bibitem[{\citenamefont{Nocedal}(1980)}]{Nocedal1980}
\bibinfo{author}{\bibfnamefont{J.}~\bibnamefont{Nocedal}},
  \bibinfo{journal}{Mathematics of Computation} \textbf{\bibinfo{volume}{35}},
  \bibinfo{pages}{773} (\bibinfo{year}{1980}), ISSN \bibinfo{issn}{0025-5718},
  \urlprefix\url{http://www.ams.org/mcom/1980-35-151/S0025-5718-1980-0572855-7/}.

\bibitem[{\citenamefont{Cao et~al.}(2014)\citenamefont{Cao, Lin, and
  Park}}]{Cao}
\bibinfo{author}{\bibfnamefont{P.}~\bibnamefont{Cao}},
  \bibinfo{author}{\bibfnamefont{X.}~\bibnamefont{Lin}}, \bibnamefont{and}
  \bibinfo{author}{\bibfnamefont{H.~S.} \bibnamefont{Park}},
  \bibinfo{journal}{Journal of the Mechanics and Physics of Solids}
  \textbf{\bibinfo{volume}{68}}, \bibinfo{pages}{239} (\bibinfo{year}{2014}),
  ISSN \bibinfo{issn}{00225096},
  \urlprefix\url{http://linkinghub.elsevier.com/retrieve/pii/S0022509614000647}.

\bibitem[{\citenamefont{Keim and Arratia}(2014)}]{Keim2014}
\bibinfo{author}{\bibfnamefont{N.~C.} \bibnamefont{Keim}} \bibnamefont{and}
  \bibinfo{author}{\bibfnamefont{P.~E.} \bibnamefont{Arratia}},
  \bibinfo{journal}{Physical Review Letters} \textbf{\bibinfo{volume}{112}},
  \bibinfo{pages}{028302} (\bibinfo{year}{2014}), ISSN
  \bibinfo{issn}{00319007},
  \urlprefix\url{http://link.aps.org/doi/10.1103/PhysRevLett.112.028302}.

\bibitem[{\citenamefont{Perchikov and Bouchbinder}(2014)}]{Perchikov2014}
\bibinfo{author}{\bibfnamefont{N.}~\bibnamefont{Perchikov}} \bibnamefont{and}
  \bibinfo{author}{\bibfnamefont{E.}~\bibnamefont{Bouchbinder}},
  \bibinfo{journal}{Physical Review E - Statistical, Nonlinear, and Soft Matter
  Physics} \textbf{\bibinfo{volume}{89}}, \bibinfo{pages}{062307}
  (\bibinfo{year}{2014}), ISSN \bibinfo{issn}{15502376},
  \urlprefix\url{http://link.aps.org/doi/10.1103/PhysRevE.89.062307}.

\bibitem[{\citenamefont{Chen et~al.}(2011)\citenamefont{Chen, Manning, Yunker,
  Ellenbroek, Zhang, Liu, and Yodh}}]{Chen2011}
\bibinfo{author}{\bibfnamefont{K.}~\bibnamefont{Chen}},
  \bibinfo{author}{\bibfnamefont{M.~L.} \bibnamefont{Manning}},
  \bibinfo{author}{\bibfnamefont{P.~J.} \bibnamefont{Yunker}},
  \bibinfo{author}{\bibfnamefont{W.~G.} \bibnamefont{Ellenbroek}},
  \bibinfo{author}{\bibfnamefont{Z.}~\bibnamefont{Zhang}},
  \bibinfo{author}{\bibfnamefont{A.~J.} \bibnamefont{Liu}}, \bibnamefont{and}
  \bibinfo{author}{\bibfnamefont{a.~G.} \bibnamefont{Yodh}},
  \bibinfo{journal}{Physical Review Letters} \textbf{\bibinfo{volume}{107}},
  \bibinfo{pages}{108301} (\bibinfo{year}{2011}), ISSN
  \bibinfo{issn}{00319007},
  \urlprefix\url{http://link.aps.org/doi/10.1103/PhysRevLett.107.108301}.

\bibitem[{\citenamefont{Gartner and Lerner}(2016{\natexlab{b}})}]{Gartner2016b}
\bibinfo{author}{\bibfnamefont{L.}~\bibnamefont{Gartner}} \bibnamefont{and}
  \bibinfo{author}{\bibfnamefont{E.}~\bibnamefont{Lerner}}
  (\bibinfo{year}{2016}{\natexlab{b}}),
  \urlprefix\url{http://arxiv.org/abs/1610.03410}.

\bibitem[{\citenamefont{Cubuk et~al.}(2014)\citenamefont{Cubuk, Schoenholz,
  Rieser, Malone, Rottler, Durian, Kaxiras, and Liu}}]{Cubuk2014}
\bibinfo{author}{\bibfnamefont{E.~D.} \bibnamefont{Cubuk}},
  \bibinfo{author}{\bibfnamefont{S.~S.} \bibnamefont{Schoenholz}},
  \bibinfo{author}{\bibfnamefont{J.~M.} \bibnamefont{Rieser}},
  \bibinfo{author}{\bibfnamefont{B.~D.} \bibnamefont{Malone}},
  \bibinfo{author}{\bibfnamefont{J.}~\bibnamefont{Rottler}},
  \bibinfo{author}{\bibfnamefont{D.~J.} \bibnamefont{Durian}},
  \bibinfo{author}{\bibfnamefont{E.}~\bibnamefont{Kaxiras}}, \bibnamefont{and}
  \bibinfo{author}{\bibfnamefont{A.~J.} \bibnamefont{Liu}},
  p.~\bibinfo{pages}{4} (\bibinfo{year}{2014}),
  \urlprefix\url{http://arxiv.org/abs/1409.6820}.

\bibitem[{\citenamefont{Paulsen et~al.}(2014)\citenamefont{Paulsen, Keim, and
  Nagel}}]{Paulsen2014}
\bibinfo{author}{\bibfnamefont{J.~D.} \bibnamefont{Paulsen}},
  \bibinfo{author}{\bibfnamefont{N.~C.} \bibnamefont{Keim}}, \bibnamefont{and}
  \bibinfo{author}{\bibfnamefont{S.~R.} \bibnamefont{Nagel}},
  \bibinfo{journal}{Physical Review Letters} \textbf{\bibinfo{volume}{113}},
  \bibinfo{pages}{068301} (\bibinfo{year}{2014}), ISSN
  \bibinfo{issn}{10797114},
  \urlprefix\url{http://link.aps.org/doi/10.1103/PhysRevLett.113.068301
  http://arxiv.org/abs/1404.4117}.

\bibitem[{rat()}]{rattler_note}
\bibinfo{note}{A rattler is defined as any particle with less than three
  contacts}.

\end{thebibliography}

\end{document}